\documentclass[letterpaper, 10 pt, conference]{IEEEtran}
\IEEEoverridecommandlockouts                              % This command is only needed if 
\usepackage{float}
\usepackage{chemformula} % Formula subscripts using \ch{}
\usepackage[T1]{fontenc} % Use modern font encodings
\usepackage{graphicx}    % Adjust table fontsize
\usepackage{amsfonts}    % Mathematical notations
\usepackage{subcaption}  % For creating subfigures
\usepackage{amssymb}
\usepackage{amsfonts}
\usepackage{commath}
\usepackage{amsmath}
\usepackage{amsthm}
\usepackage{comment}
\usepackage[version=4]{mhchem}
\usepackage{algpseudocode}
\usepackage{algorithm}

\DeclareMathOperator*{\argmin}{arg\,min}

\theoremstyle{definition}
\newtheorem{theorem}{Theorem}
\theoremstyle{definition}
\newtheorem{definition}{Definition}
\theoremstyle{definition}

\theoremstyle{definition}

\theoremstyle{definition}

\theoremstyle{definition}

\title{\LARGE \bf
Towards a BMS\textsubscript{2} Design Framework: Adaptive Data-driven State-of-health Estimation for Second-Life Batteries with BIBO Stability Guarantees
}
\author{Xiaofan Cui$^{1}$,~\IEEEmembership{Member,~IEEE}, Muhammad Aadil Khan$^{1}$,~\IEEEmembership{Student Member,~IEEE}, Surinder Singh$^{3}$, \\ Ratnesh Sharma$^{3}$, and Simona Onori$^{1}$,~\IEEEmembership{Senior~Member,~IEEE}% <-this % stops a space
\thanks{
The discussion and data are presented in part by the 2024 American Control Conference, Toronto, Canada, July 2024. This article presents new advances in analysis and new theoretical proofs for our proposed adaptive estimation law. $^{1}$ Xiaofan Cui is with the department of Electrical and Computer Engineering, University of California Los Angeles, Los Angeles, CA 90095 {\tt\small cuixf@seas.ucla.edu}. He was with the department of Energy Science \& Engineering, Stanford University at the time the research was conducted.} \thanks{$^{2}$ Muhammad Aadil Khan and Simona Onori are with the department of Energy Science \& Engineering, Stanford University, CA 94305 {\tt\small maadilk@stanford.edu, sonori@stanford.edu }(corresponding author).} \thanks{$^{3}$ Surinder Singh and Ratnesh Sharma are with the Relyion Energy, Santa Clara, CA, 95054, USA.} \thanks{$^{1,2}$ The authors would like to thank StorageX Initiative within the Precourt Institute of Energy at the Stanford University for the financial support.}}
\begin{document}
\maketitle
\thispagestyle{empty}
\pagestyle{empty}
%%%%%%%%%%%%%%%%%%%%%%%%%%%%%%%%%%%%%%%%%%%%%%%%%%%%%%%%%%%%%%%%%%%%%%%%%%%%%%%%
\begin{abstract}
A key challenge that is currently hindering the widespread  use of retired electric vehicle (EV) batteries for second-life (SL) applications is the ability to accurately estimate and monitor their state of health (SOH). Second-life battery systems can be sourced from different battery packs with lack of knowledge of their historical usage. 

To tackle the in-the-field use of SL batteries, this paper introduces an online adaptive health estimation approach with guaranteed bounded-input-bounded-output (BIBO) stability. This method relies exclusively on operational data that can be accessed in real time from SL batteries.
The effectiveness of the proposed approach is shown on a laboratory aged experimental data set of retired EV batteries.  The estimator gains are dynamically adapted  to accommodate the distinct characteristics of each individual cell, making it a promising candidate for future SL battery management systems (BMS\textsubscript{2}).
\end{abstract}

%%%%%%%%%%%%%%%%%%%%%%%%%%%%%%%%%%%%%%%%%%%%%%%%%%%%%%%%%%%%%%%%%%%%%%%%%%%%%%%%
\section{Introduction}
Accelerated adoption of lithium-ion battery (LIB) technology for various applications is pushing the transition towards sustainable energy. Electric vehicles (EVs) are at the forefront of this adoption \cite{lutsey2018power,eddy2019recharging} with some OEMs announcing a complete shift towards EV production moving away from internal combustion engine vehicles \cite{pavlinek2023transition}. While this shift is much needed to address climate change concerns, naturally, this has also increased the pressure on the supply chain for the raw materials needed to manufacture LIBs \cite{jones2023electric}. To tackle the large demand of raw materials for battery manufacturing, and fully embrace a circular economy paradigm, one promising solution lies in the reuse of EV batteries once they retire from their first-life (FL). These batteries are expected to retain approximately 70-80\% of their nominal capacity, and they could be well-suited for stationary applications such as grid storage, power generation, and end-user services e.g., uninterrupted power supply \cite{dong2023cost}. 

One challenge of working with second-life (SL) batteries is to identify accurately their actual state-of-health (SOH), given that these batteries are obtained from different battery packs and have gone through different levels of usage. 
% Due to the heterogeneity of the cells in these packs, ensuring safe usage of refurbished batteries is paramount. 

For this, we further develop a battery management system dedicated for SL batteries, referred to as BMS$_2$. BMS$_2$ is based on all the major hardware components of a conventional BMS; on the software side, BMS$_2$ still performs SOX estimation e.g., state-of-charge (SOC), using standard algorithms. In this work, though, we focus on the health estimation problem of retired EV batteries. The level of degradation of retired batteries is dependent on their usage history \cite{lu2022battery}; however, at present, historical data are not publicly available for any retired batteries \cite{anna2023}. With BMS$_2$, the idea is to overcome the lack of usage history by allowing the model to adapt to new incoming data to maintain accuracy on SOH estimation.

\section{Literature Review}
\begin{comment}
\begin{figure*}[thpb]
\centering
\includegraphics[width=0.75\textwidth]{experimental_setup_schematic.png}
\caption{\small Experimental setup. (a) Schematic of the two thermocouples $Th_1$ and $Th_2$ attached on either sides of the module surface to measure the surface temperature. (b) Schematic of the experimental setup showing the battery cycler and eight modules. The modules have insulation between one another and the thermocouples can be going inserted into the sides of the modules. On the front, positive terminal is on the top, ground terminal in the middle, and negative terminal at the bottom. The workstation is used to set the current setpoint for the battery cycler, which inputs actual current into the modules, and voltage and temperature are measured at the output. The experimental setup is present inside an HVAC-controlled room located in Santa Clara, CA at Relyion Energy Inc HQs.}
\label{fig:module}
\end{figure*}
\end{comment}

Numerous works exist that use either semi-empirical methods or physics-based models for SOH estimation \cite{Hu2022}. The former class of methods performs well on simple, well-maintained laboratory conditions, but they are prone to accuracy loss when applied to on-the-field realistic conditions \cite{Hu2022}; the latter is based on coupled partial differential equations and provides the information about internal battery states.
However, physics-based models are computationally intensive with large parameter set to be identified \cite{Pozzato2021b}. Data-driven models that are capable of learning complex degradation behaviors from input data and the choice of the model can be adjusted based on the computational resources needed to run these models onboard. SOH estimation using data-driven models has used regression techniques such as Linear Regression \cite{JIANG2018754}, Support Vector Regression \cite{Wei2018}, Gaussian Process Regression \cite{Takahashi2023}. Other works have also used neural networks \cite{Zhang2014,Bhatt2021a} that are widely regarded as universal function approximators. However, to achieve good performance on neural networks, a large dataset is required and a lot of hyperparameter tuning is needed.

While these data-driven techniques have proved useful, they are limited by the kind of training data. Traditionally, the performance of these models tends to deteriorate if the incoming data is out-of-distribution \cite{li2022uncertainty} in the sense of statistically different from the kind of data that was used to train the model. Existing work around adaptive estimation utilizes periodic stream of health indicators by performing reference performance tests (RPTs) and using feedback to update the model \cite{zhang2022machine,she2021offline,zhou2013optimized,XING2013811}. Evidently, this is infeasible for BMS$_2$ since SL batteries cannot be easily removed from service and diagnosed once they are deployed \cite{Zhang2022a}. Another popular approach adopted for this problem uses transfer learning in which a baseline model is trained using a neural network. Afterwards, the initial layers of the model are frozen while the final few layers are retrained on the target domain data \cite{von2022state,zhang2023voltage}. The limitation of transfer learning resides in the target domain used to re-train the model. In field operation, the uncertainty of the incoming data is high and it is difficult to ensure that the model performs consistently.

The health estimation task aspect of the BMS$_2$ design not only requires online functionality but also demands adaptability. This requires the estimator model to dynamically adjust and evolve as more online measurements. 
In principal, it should enhance the model's generalizability to various operating conditions and unforeseen SL batteries once they are deployed. 
%Moreover, taking into account the auto-correlation of battery SOH aging as a time series \cite{severson2019data}, historical health data gains significance in the estimation of the current SOH. 
From a practical standpoint, it is essential for health estimator to guarantee the error to stay within specified bounds. 
%Nonetheless, there is a dearth of literature providing mathematical assurances for these boundaries.

%To ensure the SOH estimation algorithms trained offline, in this work, 
% The paper is organized as follows: in section 1, we introduce the paper, in section 2, we reviewed the major literature, in section 3, we go over the battery  in section 4,
In this work, we introduce an adaptive SOH estimation method that combines clustering-based estimation with regression-based estimation.
The stability of the adaptive framework is theoretically shown via bounded-input bounded-output (BIBO) stability.
%We compare the results obtained from the adaptive framework with the results obtained from the offline model to show our model adaptation improves the estimation accuracy.
The rest of this article is organized as follows: Section III presents the retired battery dataset. 
The offline data-driven estimator and online adaptive estimator are discussed in Sections IV and V, respectively.
The health estimation results on different testing scenarios are illustrated in Section VI. 
Finally, Section VII concludes this article.

\section{Retired Battery Dataset}
\begin{figure}[thbp]
\centering
\includegraphics[width=0.8\columnwidth]{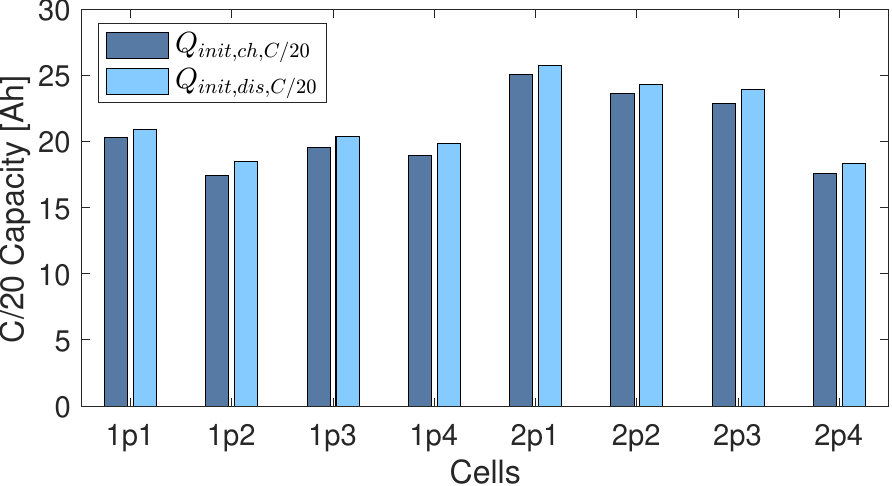}
\caption{\small Initial distribution of C/20 capacity for all SL cells with a small difference between initial charge capacity $Q_{init,ch,C/20}$ and initial discharge capacity $Q_{init,dis,C/20}$. Cell 2.1 has the highest initial capacity while Cell 2.4 has the lowest initial capacity among all cells.}
\label{fig:Q_initial}
\vspace{-5pt}
\end{figure}

The dataset used in this work consists of eight retired pouch cells obtained from Nissan leaf EV battery packs with LMO/graphite chemistry. The nominal capacity (fresh cell) is 33.1\,Ah and the voltage range is from 2.5\,V to 4.2\,V. 
%The cells are arranged in modules where each module contains four pouch cells in a 2s2p configuration accessible via the three terminals outside the module. Based on the module design, a pair of terminals (must include the middle terminal) can be used to access 2p cells while positive and negative terminal together can be used to access the complete 2s2p configuration. In this work, we access only the 2p cells using the positive and middle terminals and refer to these 2p cells as a single cell for the remainder of this paper. Hence, the eight cells in the dataset are, in essence, eight 2p cells. Surface temperature of the cells is measured by a pair of thermocouples $Th_1$ and $Th_2$ attached to the side of all the modules. 
%The complete experimental setup is shown in Fig.\,\ref{fig:module}.
%Since these are retired cells with an unknown usage history, it is important to characterize these cells to understand their existing SOH.
An experimental campaign is designed which consists of three different reference performance tests (RPTs) and an aging profile repeated multiple times in between RPTs. The cycling profile is designed to mimic the load expected in a grid-storage application in a simplistic manner \cite{cui2023}. 
The RPTs consists of C/20 capacity test, a Hybrid Pulse Power Characterization (HPPC) test, and a C/40 Open-circuit voltage (OCV) test \cite{Ha2024}. One important aspect of this experimental campaign is that, except for the OCV test performed between 2.5 and 4.2V, all the other tests are performed between 3V to 4V. This kind of voltage derating is well-suited for SL applications to guarantee safe operation by limiting the growth of solid electrolyte interphase (SEI) layer and thereby, reducing battery degradation \cite{aiken2022li}. The initial distribution of C/20 capacity of these cells is shown in Fig.\,\ref{fig:Q_initial}. $Q_{init,ch,C/20}$ is the initial C/20 charge capacity and $Q_{init,dis,C/20}$ is the initial C/20 discharge capacity obtained from the charge and discharge portions of the C/20 test, respectively. A small difference between $Q_{init,ch,C/20}$ and $Q_{init,dis,C/20}$ indicates that these cells have high coulombic efficiency. Detailed information about the experimental campaign, including length of experimental campaign and design of testing protocols, is given in \cite{cui2023}.
\begin{figure}[thpb]
\centering
\includegraphics[width=0.85\columnwidth]{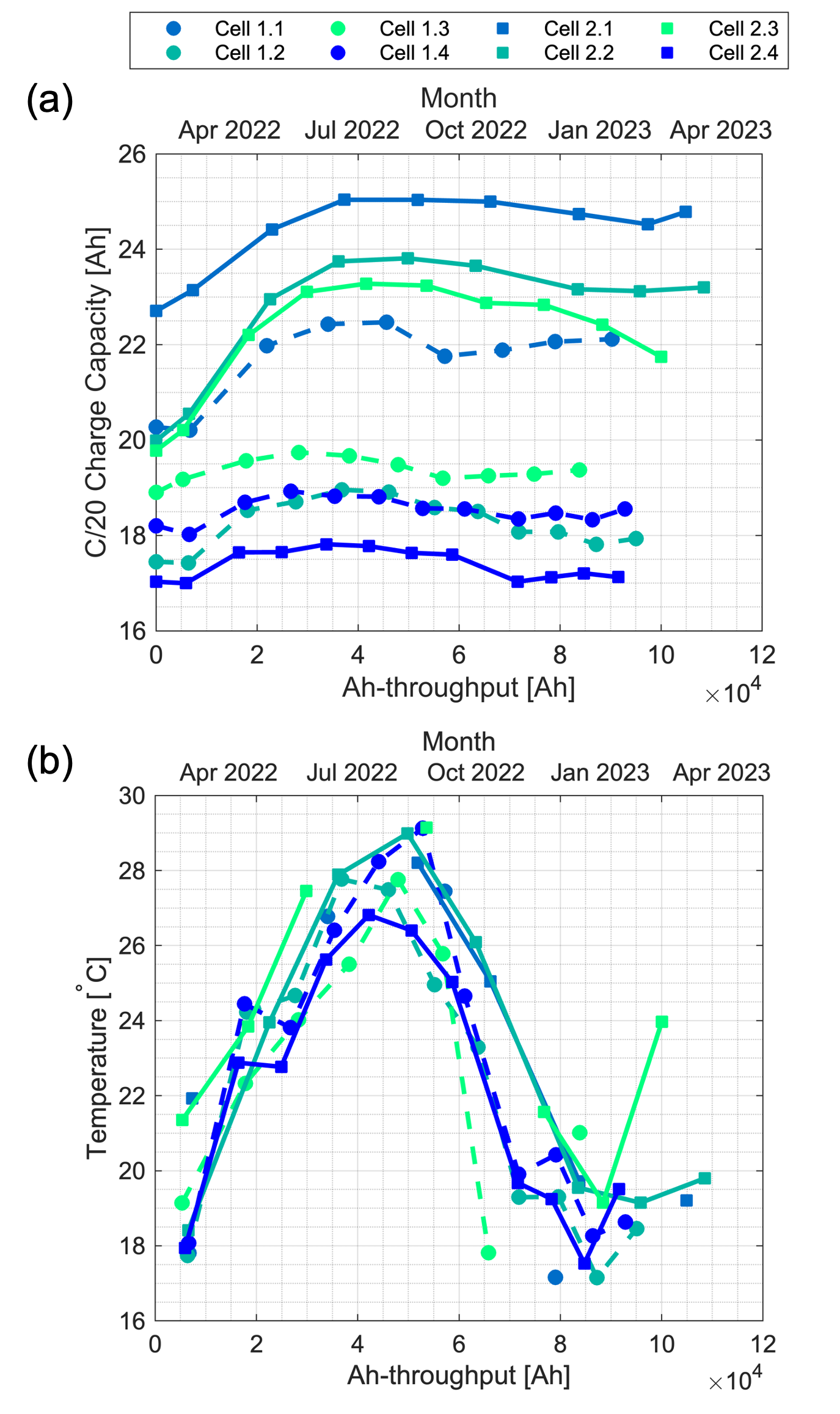}
\caption{\small Capacity and temperature for the SL cells used in this work. (a) C/20 charge capacity as a function of Ah-throughput. (b) Temperature as a function of Ah-throughput (below) and months (above) with a parabolic shape reaching a maximum value during the months of August/September 2022 indicating the effect of seasonal variation of temperature on all the cells. Gaps in temperature lines represent missing data for those cells e.g., Cell 2.3.}
\label{fig:capacity_temperature}
\vspace{-10pt}
\end{figure}

%Fig.\,\ref{fig:capacity_temperature}(a) \textcolor{red}{shows that charge capacity of these cells obtained at a C-rate of C/20}, it can be seen that Cell 2.4 is the most aged cell since it has the lowest initial capacity at the start of SL. On the other hand, Cell 2.1 is the least aged cell while all the other cells fall in between these two. The cells are present inside a Heating, Ventilation and Cooling (HVAC) controlled room. Although the temperature of the room is maintained at a human-comfort level, the seasonal variations in temperature influence the cell surface temperature as shown in Fig.\,\ref{fig:capacity_temperature}(b). For each cell, as shown in Fig.\,\ref{fig:module}(c), two thermocouples are attached to the surface of the cells resulting in a total of 16 thermocouples. 

The C/20 charge capacity trajectory is shown in Fig.\,\ref{fig:capacity_temperature}(a). As opposed to conventional battery degradation profiles \cite{attia2022knees}, these cells exhibit an increase in capacity with all cells reaching a maximum capacity point which is higher than their initial capacities. As outlined in detail in \cite{cui2023}, this behavior is due to the varying temperature that the cells experience during the experimental campaign (see Fig.\,\ref{fig:capacity_temperature}(b)). The dataset, being the first of its kind, highlights how cells can potentially behave when used in practical SL applications, especially in cases where the temperature cannot be controlled. Furthermore, the effect of voltage derating can be observed by the fact that cells, upon reaching the one-year mark from start of testing, still possess higher or equal capacity compared to their initial capacities. 

\section{Offline Data-driven Model} 
The following notation is used in the paper:
\begin{comment}
\begin{itemize}
    \item[(1)] Sequence $\{a_n\}_{n\le N} \triangleq \{a_1, a_2, \cdots, a_N\}$ where $a_n$ represents the general item of the sequence $\{a_n\}$, and $N$ represents the length of the sequence. For ease of notation, if not specified, $\{a_n\}$ is equivalent to $\{a_n\}_{n\le N}$.
    \item[(2)] $\|\cdot\|_2$ represents the $\mathcal{L}^{2}$ norm. The $\mathcal{L}^{2}$ norm of a sequence $\{x_n\} = (x_1, x_2, \cdots, x_N)$ is captured by the formula $\| \{x_n\} \|_2 = \sqrt{\sum_{n = 1}^{N} x_n^2}$.
    \item[(3)] $\|\cdot\|$ represents the $\mathcal{L}^{\infty}$ norm. The $\mathcal{L}^{\infty}$ norm of a sequence $\{x_n\} = (x_1, x_2, \cdots, x_N)$ is captured by the formula $\| \{x_n\} \|_{\infty} = \max_{n = 1}^{N} \left| x_n \right|$.
    \item[(4)] $\mathbb{R}^{+} = \{z \in \mathbb{R}: z > 0 \}$.
    \item[(5)] Three performance metrics include Mean Absolute Percentage Error (MAPE), Root Mean Squared Error (RMSE), and Root Mean Squared Percentage Error (RMSPE), which are defined by
    \begin{gather}
        \text{MAPE} = \frac{1}{M} \sum_{y \in Y,\;\hat{y} \in \hat{Y}}\frac{\left|\hat{y} - y\right|}{y} \times 100\,\% \\
        \text{RMSE} = \sqrt{\frac{1}{M} \sum_{y \in Y,\;\hat{y} \in \hat{Y}} \left(\hat{y} - y\right)^2} \times 100\,\% \\
        \text{RMSPE} = \sqrt{\frac{1}{M} \sum_{y \in Y,\;\hat{y} \in \hat{Y}} \left(\frac{\hat{y} - y}{y}\right)^2} \times 100\,\% 
    \end{gather}
    where $Y$ is the measured data, $\hat{Y}$ is the model estimation, and $M$ is the number of samples.
\end{itemize}
\end{comment}
Sequence $\{a_n\}_{n \le N} \triangleq \{a_1, a_2, \cdots, a_N\}$ where $a_n$ represents the general term of the sequence $\{a_n\}$, and $N$ represents the length of the sequence. For ease of notation, if not specified, $\{a_n\}$ is equivalent to $\{a_n\}_{n \le N}$.
$\|\cdot\|_2$ represents the $\mathcal{L}^2$ norm. The $\mathcal{L}^2$ norm of a sequence $\{x_n\} = (x_1, x_2, \cdots, x_N)$ is defined by the formula $\| \{x_n\} \|_2 = \sqrt{\sum_{n=1}^N x_n^2}$.
$\|\cdot\|_{\infty}$ represents the $\mathcal{L}^{\infty}$ norm. The $\mathcal{L}^{\infty}$ norm of a sequence $\{x_n\} = (x_1, x_2, \cdots, x_N)$ is defined by the formula $\| \{x_n\} \|_{\infty} = \max_{n=1}^N \left| x_n \right|$.
$\mathbb{R}^{+} = \{z \in \mathbb{R} : z > 0 \}$.
Three performance metrics include Mean Absolute Percentage Error (MAPE), Root Mean Squared Error (RMSE), and Root Mean Squared Percentage Error (RMSPE), which are defined as
\begin{gather}
    \text{MAPE} = \frac{1}{M} \sum_{y \in Y, \; \hat{y} \in \hat{Y}} \frac{\left| \hat{y} - y \right|}{y} \times 100\,\% 
\end{gather}
\begin{gather}
    \text{RMSE} = \sqrt{\frac{1}{M} \sum_{y \in Y, \; \hat{y} \in \hat{Y}} \left( \hat{y} - y \right)^2} 
\end{gather}
\begin{gather}
    \text{RMSPE} = \sqrt{\frac{1}{M} \sum_{y \in Y, \; \hat{y} \in \hat{Y}} \left( \frac{\hat{y} - y}{y} \right)^2} \times 100\,\%
\end{gather}
where $Y$ is the measured data, $\hat{Y}$ is the model estimation, and $M$ is the number of samples.

The SOH for SL batteries is the current C/20 capacity normalized by their respective initial C/20 capacity at the start of SL.
\footnote{Other indicators of SOH, such as State of Energy (SOE), could potentially serve as alternative candidates. \cite{Takahashi2023} has suggested a high correlation between SOE and SOH for second-life batteries.}
% Therefore, in this paper, we choose to utilize C/20 capacity to calculate the SOH indicator.
For this work, an Elastic-Net Regression (ENR) model is selected which provides ease of training and adaptability as well as low computational requirements. The choice of data-driven model is dictated by the complexity of the data, computational resources, and the training time required to train these models. In our case, another factor that is taken into account is the ease of parameter adaptation when new data becomes available. 
%Furthermore, since our dataset is small in terms of the number of cells, this model provides a suitable framework for the health estimation problem \cite{cui2023}. 
In ENR, the loss function used to train the model consists of a linear combination of $L_1$ (lasso) and $L_2$ (ridge) regularization terms. The following optimization problem is solved to train the model:
\begin{gather}
	\hat{\beta} = \argmin_{\beta_0, \beta, \lambda} \,\, \|Y - X \beta - \beta_0 \mathbf{1}_{n\times 1}\|_2 + \nonumber \\
	\lambda \left(1-\alpha \right)\|\beta\|_2^2 + \lambda \alpha \|\beta\|_1
\end{gather}
where $Y \in \mathbb{R}^n$ represents the SOH indicators, $X \in \mathbb{R}^{n \times m}$ contains $m$ features each with $n$ observations, $\beta \in \mathbb{R}^m$ contains $m$ regression coefficients, and $\beta_0, \lambda, \alpha \in \mathbb{R}$ are the scaler intercept, the hyperparameter to adjust the influence of $L_1$ and $L_2$, and the regularization parameter, respectively. 
\begin{figure}[thpb]
\centering
\includegraphics[width=0.85\columnwidth]{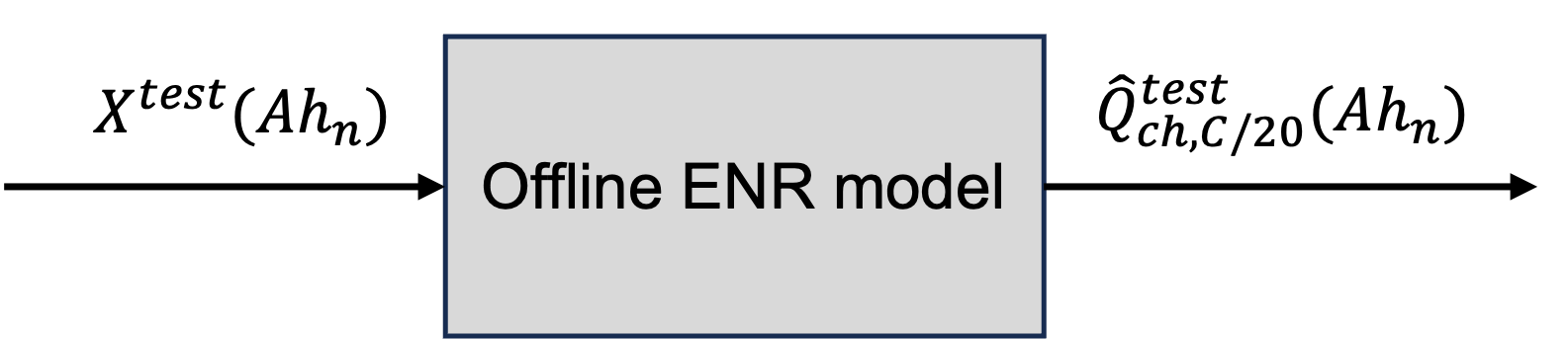}
\caption{\small ENR model with input features $X^{test}$ from the test set and an estimated scalar output $\hat{Q}^{test}_{ch,C/20}$ as a function of Ah-throughput.}
\label{fig:offline_enr}
\end{figure}

%\subsection{Offline ENR}
% The output is chosen to be the C/20 charge capacity. 
%It is noted that due to the high coulombic efficiency of the cells, the difference between charge and discharge capacity is minimal. 
Fig.\,\ref{fig:offline_enr} shows the flow of data through the offline ENR model where $X^{test}(Ah_n)$ represents the input features of the test set at $Ah_n$ and $\hat{Q}^{test}_{ch,C/20}(Ah_n)$ is the estimated C/20 charge capacity. More details about the offline model are given in \cite{cui2023}. 
\begin{comment}
From a total of 8 cells, 6 cells are used for training and 2 cells are used for testing. The model is referred to as ``offline'' because the estimation results are based purely on the existing training data and no attempt is made to adapt the model to improve its estimation based on new data. From a total of 8 cells, the number of possible combinations of 2 cells without repetition is $_8C_2=28$. %given by 
% \begin{gather}
% 	_nC_r = \frac{n!}{r! (n-r)!}
% \end{gather}
\end{comment}
\begin{figure*}[h]
    \centering
\includegraphics[width=\textwidth]{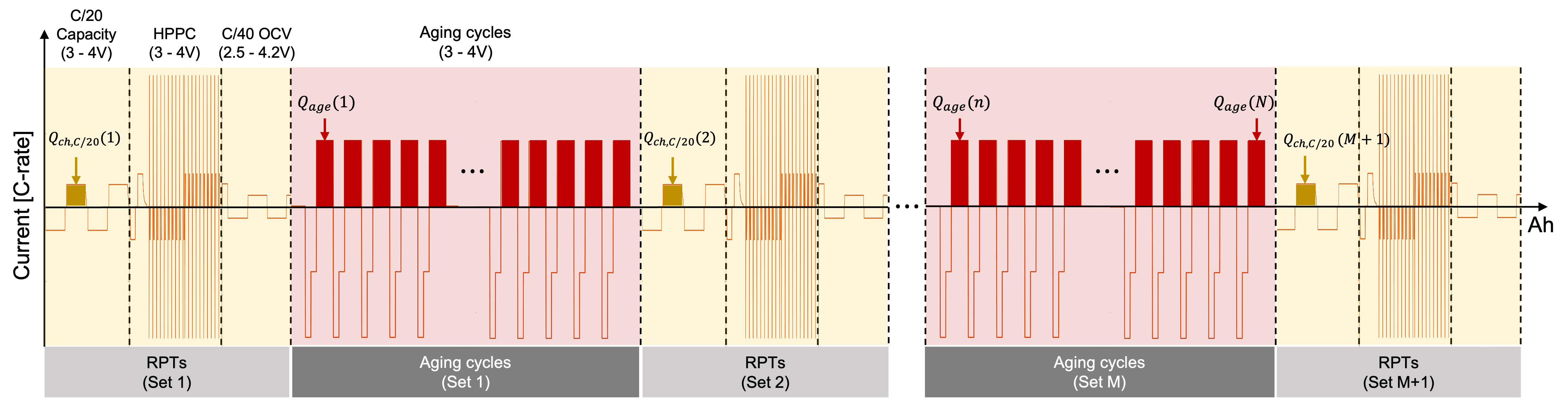}
    \caption{\small Design of experiments showing Reference performance tests (RPTs) and aging cycles. RPTs contain C/20 capacity test, HPPC test, and OCV test, which makes one set. For the complete campaign, $N$ sets of aging cycles are performed while $M+1$ sets of RPTs are performed. $N >> M$ for the dataset. $Q_{ch,C/20}$, the battery health indicator, is obtained once from the C/20 capacity test of each RPT set, as highlighted in brown. $Q_{age}$, the aging cycle charge, is obtained from each cycle, as highlighted in red. The limits of $Q_{ch,C/20}(i)$ are given by $t_{ch}^{C/20}(i)$ and $t_{dis}^{C/20}(i)$ where $i=1,2,...,M+1$, and the limits of $Q_{age}(j)$ are given by $t_{ch}^{age}(j)$ and $t_{dis}^{age}(j)$ where $j=1,2,...,n,...,N$. Positive current indicates charge and negative current indicates discharge.}
    \label{fig:df12_plots}
\end{figure*}
\section{Online Adaptive Estimator}
The health estimation of BMS\textsubscript{2} not only needs to be online but also adaptive, meaning that the estimator model should adapt and evolve as new online measurements are fed into the BMS\textsubscript{2}.

The aging-cycle charge throughput $Q_{age}$ is calculated as
\begin{align}
    Q_{age}(n) = \int_{t^{age}_{ch}(n)}^{t^{age}_{dis}(n)} I(t)\,dt,
\end{align}
where $Q_{age}(n)$, $t_{ch}^{age}(n)$, $t_{dis}^{age}(n)$, and $I(t)$ are shown in Fig.\,\ref{fig:df12_plots}. The C/20 charge capacity $Q_{ch, C/20}$ is calculated as
\begin{align}
    Q_{ch,C/20}(n) = \int_{t^{C/20}_{ch}(n)}^{t^{C/20}_{dis}(n)} I(t)\,dt,
\end{align}
where $Q_{ch, C/20}(n)$, $t^{C/20}_{ch}(n)$, $t^{C/20}_{dis}(n)$, and $I(t)$ are shown in Fig.\,\ref{fig:df12_plots}.
The accumulated Ah throughput for the $n^\text{th}$ aging cycle $Ah_n^{age}$ is calculated as
\begin{align}
    Ah_n^{age} = \int_{0}^{t_{ch}^{age}} \Big|I(t)\Big|\,dt.
\end{align}
The accumulated Ah throughput for the $n^\text{th}$ C/20 charge-discharge cycle $Ah_n^{ch, C/20}$ is calculated as
\begin{align}
    Ah_n^{ch, C/20} = \int_{0}^{t_{ch}^{C/20}} \Big|I(t)\Big|\,dt.
\end{align}
\begin{definition} \label{def: q_aging_traj}
  Given the monotonically increasing accumulated Ah-throughput sequence $\{Ah^{age}_n\} = \left(Ah^{age}_1, Ah^{age}_2, \cdots, Ah^{age}_N\right)$ and the aging-cycle charge throughput sequence $\{Q_{age}\} =  \left(Q_{age}(1), Q_{age}(2), \cdots, Q_{age}(N)\right)$, then the aging-cycle charge throughput trajectory is a curve denoted by $Q_{age}(\{Ah^{age}_n\})$ or simply $Q_{age}$.
\end{definition}
% \emph{Remark}:
% While $Q_{age}$ is one of the numerous online-available features, it stands out as the most relevant feature with respect to our SOH indicator $Q_{}$ \cite{cui2023}. Consequently, our adaptive estimation algorithm utilizes this feature to update the model.
\begin{definition} \label{def: capacity_traj}
  Given the monotonically increasing accumulated Ah-throughput sequence $\{Ah^{}_n\} = \left(Ah^{ch, c/20}_1, Ah^{ch, c/20}_2, \cdots, Ah^{ch, c/20}_N\right)$
 and the C/20 charge capacity sequence $\{Q_{ch,C/20}\} =  \left(Q_{ch,C/20}(1), Q_{ch,C/20}(2), \cdots, Q_{ch,C/20}(N)\right)$, then the C/20 charge capacity trajectory is a curve denoted by $Q_{ch, C/20}(\{Ah^{ch, c/20}_n\})$ or simply $Q_{ch, C/20}$.
\end{definition}
\emph{Remark}: $Q_{ch, C/20}$ is calculated as shown in Fig.\,\ref{fig:df12_plots}. For ease of notation, in the following derivations,
the C/20 charge capacity sequence is simply denoted by $\{Q_{n}\}$ and 
$Q_{ch, C/20}$ is simply denoted by $Q$.

The \emph{normalized SOH indicator} is the normalized C/20 charge capacity, calculated by 
\begin{align} \label{eqn:norm_q_ch}
    \bar{Q}(Ah_n) = \frac{Q_{}(Ah_n)}{Q_{}(0)},
\end{align}
where $Q_{}(0)$ is the $C/20$ charge capacity of the cell measured at the beginning of the second-life experiment. For ease of notation, in the following derivations of this section, $Q_{}(0)$ is simply denoted by $Q_{0}$.
% by following the similar definition in (\ref{def: capacity_traj}), the \emph {normalized SOH indicator trajectory} can be defined.
\begin{definition} \label{def:train_set_notation}
    The $K$ cells in the \emph{training set} are denoted by indices $1, \cdots, K$. Therefore, the \emph{C/20 charge capacity trajectories} in the training set are $Q^{1}_{}, \cdots, Q^{K}_{}$, denoted by $Q^{train}_{}$. The \emph{aging cycle charge throughput trajectories} in the training set are $Q^{1}_{age}, \cdots, Q^{K}_{age}$, denoted by $Q^{train}_{age}$. Other than the aging cycle charge throughput trajectories, there are many other feature trajectories in the training set. We denote all feature trajectories in the training set by $X^{train}_{age}$.
\end{definition}
\begin{definition} \label{def:test_set_notation}
    The one cell in the \emph{test set} is denoted by index $z$. Therefore, the \emph{C/20 charge capacity trajectory} in the test set is $Q^{z}_{}$, denoted by $Q^{test}_{}$. The \emph{aging cycle charge throughput trajectory} in the test set is $Q^{z}_{age}$, denoted by $Q^{test}_{age}$. Other than the aging cycle charge throughput trajectories, there are many other feature trajectories in the test set. We denote all feature trajectories in the test set by $X^{test}_{age}$.
\end{definition}
\emph{Remark}: In this dataset, Cells 1.1, 1.2, 1.3, 2.1, 2.2, 2.3, 2.4 correspond to $k$ = 1, 2, 3, 4, 5, 6, 7, respectively. Cell 1.4 used for testing corresponds to $z$.
% \begin{definition}
% The training set of the \emph{aging-cycle charge trajectories} is denoted by $Q_{age}^{1}, \cdots, Q_{age}^{K}$.
% %where $K$, the number of cells, is the cardinality of the set.
% \end{definition}
%Without loss of generality, in the following,  battery cells 1 to $K$ are used as the training set and cell $z$ as the testing set.
\begin{comment}
\begin{definition} \label{def:dd_est}
    The \emph{data-driven estimator} is a mapping $\mathcal{E}$: 
    \begin{align}
        \hat{y}^{test}_n & =  \mathcal{E} \left(X^{test}_n; \{X^{train}_n\}, \{y^{train}_n\}\right) \\
        e_n & =  \hat{y}^{test}_n - y^{test}_n
    \end{align}
    where $\{y^{train}_n\}$ denotes the training response trajectory set;
    $y^{test}_n \in \mathbb{R}^N $ and $y^{test}_n  \in \mathbb{R}^N $ denote the true and estimated test response trajectories respectively;
    $e_n$ represents the estimation error trajectory;
    $\{X^{train}_n\}$ denotes the training feature trajectory set; 
    $X^{test}_n$ denotes the test feature trajectory.
\end{definition} 
\end{comment}

\begin{definition} \label{def:dd_soh_est}
    The \emph{battery health estimator} $\mathcal{E}_{H}$ is a type of data-driven estimator of the form: 
    \begin{align}
        \hat{Q}_{}^{test} & = f\left( X^{test}_{age}, X^{train}_{age}, Q_{}^{train} \right),
    \end{align}
    where $f$ is a generic mapping function, $\hat{Q}_{}^{test}$ represents the estimated C/20 charge capacity of the cell in the test set; $Q^{train}_{}$ and $X^{train}_{age}$ can be found in Definition\,\ref{def:train_set_notation};
    $X^{test}_{age}$ can be found in Definition\,\ref{def:test_set_notation}.
\end{definition}
% \emph{Remark}: The $\{\cdot\}$ represents the sequence of observation in the training set. The single-observed training response trajectory $y^{train}_n \in \mathbb{R}^N$. Given single feature is used, the single-observed training feature trajectory $X^{train}_n \in \mathbb{R}^N$. Given M different features are used, the single-observed training feature trajectory $X^{train}_n \in \mathbb{R}^{N\times M}$.
% \emph{Remark}: The $\{\cdot\}$ represents the sequence of observation in the training set. The single-observed training response trajectory $y^{train}_n \in \mathbb{R}^N$. Tje feature trajectory $Q^{train}_{aging} \in \mathbb{R}^N$.
%Given M different features are used, the single-observed training feature trajectory $X^{train}_n \in \mathbb{R}^{N\times M}$.
%\vspace{+3 pt}
The estimation error $e_n$ is evaluated as
\begin{align}
    e & =  \hat{Q}^{test} - Q^{test},
\end{align}
where $Q^{test}$ can be found in Definition\,\ref{def:test_set_notation}.
    
Stability is of practical importance for adaptive estimation laws as it ensures that the proposed estimation process does not result in divergence. This paper uses the notion of bounded-input, bounded-output (BIBO) stability as follows:
\begin{definition} \label{eqn:bibo_dde}
    An adaptive estimator $\mathcal{E}_H$ defined in Definition \ref{def:dd_soh_est} is \emph{BIBO stable} if for any bounded input signal with $\|X^{train}_{age}\| < \infty$,  $\|X^{test}_{age}\| < \infty$,  $\|Q^{train}_{}\| < \infty$, $\| Q^{test}_{} \| < \infty$, the error $e_n$ satisfies $\|e_n\| < \infty$.
\end{definition}

\subsection{Clustering-based adaptive estimation}
The clustering-based adaptive estimation algorithm aims to find the trajectory in the training set that is closest to the trajectory in the test set. The closeness metric is defined by the following distance function:
\begin{definition} \label{def:dist_func}
The \emph{distance} between two aging-cycle charge throughput trajectories $Q^x_{age}$ and $Q^y_{age}$ is defined as
\begin{align}
    \text{dist}(Q^x_{age}, Q^y_{age}) = \sqrt{\sum_{i = 1}^{n} \left(Q^x_{age}(Ah_i) - Q^y_{age}(Ah_i)\right)^2}, & \nonumber \\
    1 \le n \le N. &
\end{align}
\end{definition}
\begin{algorithm*}
\caption{\footnotesize Calculate Distance}
\label{alg:calculate_distance}
\footnotesize
\begin{algorithmic}
\Function{CalculateDistance}{$Q^x_{\text{age}}\left(Ah_1, \cdots, Ah_N \right)$, $Q^y_{\text{age}}\left(Ah_1, \cdots, Ah_N \right)$}
    \State $\text{distance} \gets 0$ \Comment{Initialize distance to zero}
    \For{$i$ from $1$ to $n$}
        \State $\text{diff} \gets Q^x_{\text{age}}(Ah_i) - Q^y_{\text{age}}(Ah_i)$
        \State $\text{distance} \gets \text{distance} + \text{diff}^2$
    \EndFor
    \State $\text{distance} \gets \sqrt{\text{distance}}$ \Comment{Calculate the square root}
    \State \Return $\text{distance}$ \Comment{Return the distance}
\EndFunction
\end{algorithmic}
\end{algorithm*}
% \begin{definition} \label{def:dist_func}
% The \emph{distance} between two aging-cycle charge throughput trajectories $Q^x(\{Ah_i\}_{i \le n})$ and $Q^y(\{Ah_i\}_{i \le n})$ are 
% \begin{align}
%     \text{dist}(Q^x, Q^y) = \Big| Q^x(Ah_n) - Q^y(Ah_n) \Big|,
% \end{align}
% \end{definition}
\emph{Remark}: 
The pseudo-code to calculate the distance can be found in Algorithm\,\ref{alg:calculate_distance}.
$\text{dist}(Q^x_{age}, Q^y_{age})$ is non-negative and symmetric quantity.
Compared to the nearest neighbor distance metric based on the latest measurements, this trajectory distance metric memorizes all historical differences between $Q^x_{age}$ and $Q^y_{age}$ and evaluates similarities based not only on instantaneous measurements but also on historical capacity measurements. Therefore, it is more robust in the presence of one-shot measurement noise.
One shortcoming of this distance metric is that it requires more than one data point to make a definitive classification. However, in the long run, when the online BMS\textsubscript{2} has accumulated enough information, this additional data requirement can be easily met.
This distance function can be easily generalized to measure the distance between other trajectories such as the C/20 charge capacity trajectory, etc.

\begin{figure}[htbp]
    \centering
    \includegraphics[width=0.9\columnwidth]{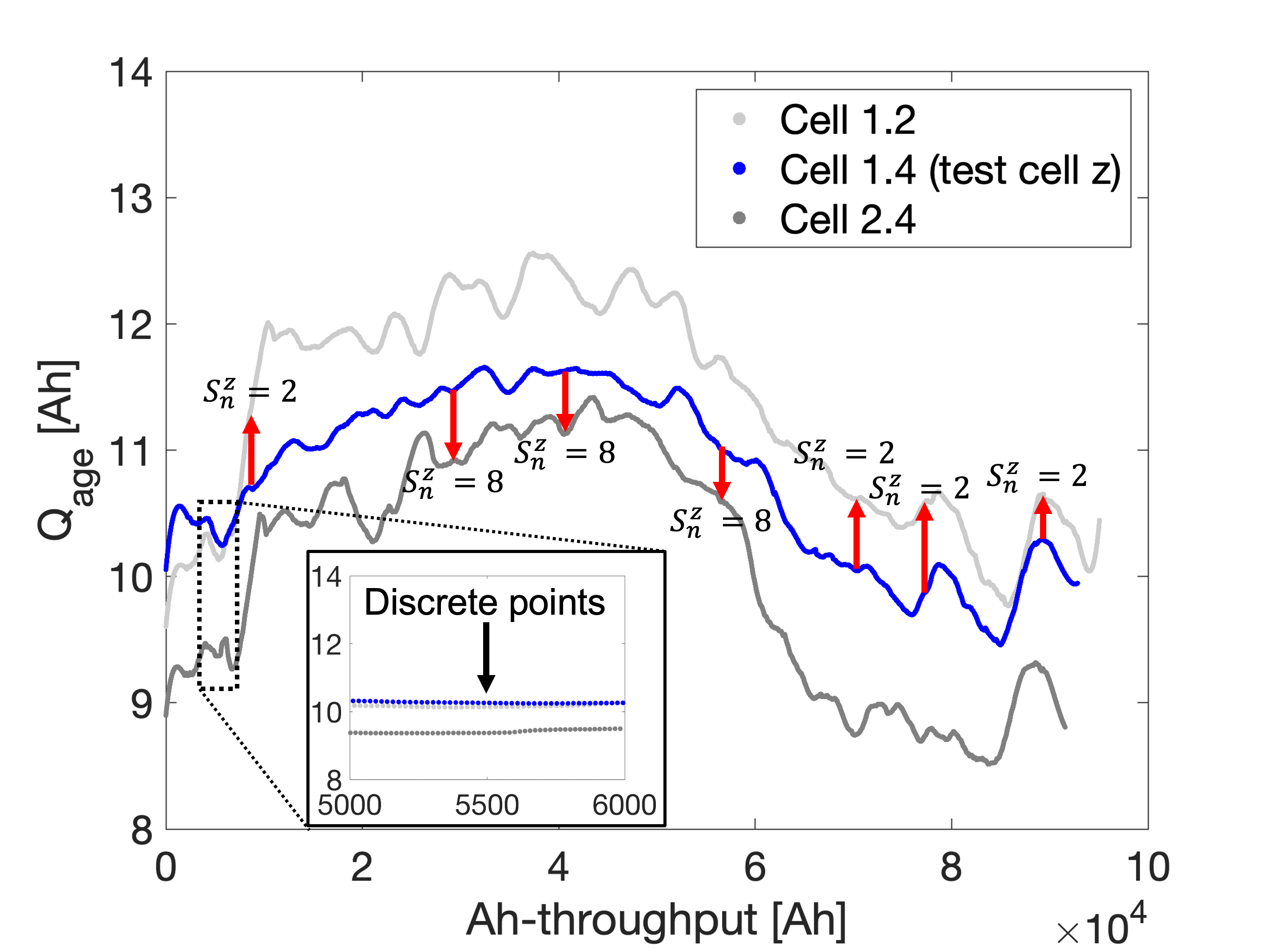}
    \caption{\small Example of cell classification defined in Definition \ref{eqn:class_ind_seq} with Cell 1.4 as the test cell compared against Cell 1.2 ($S_n^z=2$) and Cell 2.4 ($S_n^z=8$). Red arrows points to the aging cycle charge trajectory with the minimum distance from $Q_{age}^z$ -- where $z=1.4$ -- at different Ah-throughputs. Zoomed plot shows that these are highly dense discrete points since $Q_{age}^z$ is extracted from every cycle.}
    \label{fig:cell_classification}
    \vspace{-20pt}
\end{figure}

Whenever a new sample of cell $z$ is collected, the classification is updated, giving the clustering index that cell $z$ is assumed to belong to:
\begin{definition} \label{eqn:class_ind_seq}
     \emph{Classification index sequence} for cell $z$, $\{S^{z}_{n}\} = \left( S^{z}_{1}, S^{z}_{2}, \cdots, S^{z}_{N} \right)$ is a discrete-time sequence defined by: %mapping: $\mathbb{Z}^{+} \rightarrow \mathbb{R^{+}}$:
\end{definition}
% The Clustering algorithm figures out the mapping between the clustering index sequence
\begin{align}\label{eqn:clu_ind_seq}
    S^{z}_{n} = & \underset{1 \leq k \leq K}{\text{argmin}}\, \text{dist}\left(Q^{z}_{age}(\{Ah_i\}_{i \le n}), Q^{k}_{age}(\{Ah_i\}_{i \le n}) \right),
\end{align}
where $n \in \left[1, 2, \cdots, N\right]$, $Q^{z}_{age}$ and $Q^{k}_{age}$ are 
the aging-cycle charge trajectories of cell $z$ and $k$, respectively. The distance function $dist$ has been defined in Definition \ref{def:dist_func}. An example of the cell classification is shown in Fig.\,\ref{fig:cell_classification}. The pseudo-code to calculate the classification index sequence can be found in Algorithm\,\ref{alg:classification_index_sequence}.
\begin{algorithm*}
\caption{\footnotesize Classification Index Sequence}
\label{alg:classification_index_sequence}
\begin{algorithmic}
\footnotesize
\Function{ClassificationIndexSequence}{$Q^z_{\text{age}}\left(Ah_1, \cdots, Ah_N \right)$, 
$Q^1_{\text{age}}\left(Ah_1, \cdots, Ah_N \right)$, $\cdots$, $Q^K_{\text{age}}\left(Ah_1, \cdots, Ah_N \right)$}
    \State Initialize an empty sequence $S^z$
    \For{$n$ from $1$ to $N$}
        \State $\text{minDist} \gets \infty$ \Comment{Initialize minimum distance to positive infinity}
        \State $S_n^z \gets 0$ \Comment{Initialize the classification index for the current time step}
        \For{$k$ from $1$ to $K$}
            \State $\text{dist} \gets$ \Call{CalculateDistance}{$Q^z_{\text{age}}\left(Ah_1, Ah_2, \cdots, Ah_n \right), Q^k_{\text{age}}\left(Ah_1, Ah_2, \cdots, Ah_n \right)$}
            \If{$\text{dist} < \text{minDist}$}
                \State $\text{minDist} \gets \text{dist}$ \Comment{Update minimum distance}
                \State $S_n^z \gets k$ \Comment{Update the classification index}
            \EndIf
        \EndFor
        \State Append $S_n^z$ to $S^z$ \Comment{Add the classification index to the sequence}
    \EndFor
    \State \Return $S^z$ \Comment{Return the classification index sequence}
\EndFunction
\end{algorithmic}
\end{algorithm*}

\begin{theorem} \label{theorem:ddm_stable}
Consider a data-driven estimator described in Definition\,\ref{def:dd_soh_est}. Let the  
battery cells 1, $\cdots$, $K$ be the training set and cell $z$ be the test set. If the model parameters $\lambda_k$ are adapted according to
\begin{align} \label{eqn:lambda_adaptation}
    \lambda_k = \frac{\sum_{n = 1}^{N}   \resizebox{!}{1.5\height}{$\chi$}(S^z_n = k) Ah_n}{\sum_{n = 1}^{N} Ah_n},
\end{align}
where 
\begin{align}
    \resizebox{!}{1.5\height}{$\chi$}(S^z_n = k) = \begin{cases}
        1 & \text{if} \,\, S^z_n = k, 1 \le k \le K \\
        0 & \text{otherwise},
    \end{cases}
\end{align}
$S^z_n$ is defined in (\ref{eqn:clu_ind_seq}). Then the adaptive estimator, formulated as
\begin{align}
    \hat{Q}^{z}_{}(Ah_n) &= Q^{z}_{0}\hat{\bar{Q}}^z(Ah_n), \label{eqn:norm_q_z} \\
    \hat{\bar{Q}}^{z}(Ah_n) &= \sum_{k = 1}^{K} \lambda_k
    \bar{Q}^k(Ah_n) \label{eqn:norm_q_z_decomp},
\end{align}
where $\hat{Q}^{z}_{}$, the estimation target, represents the estimated C/20 charge capacity of cell $z$, is BIBO stable.
%, i.e. the capacity estimates converge to a bounded error as $Ah \rightarrow \inf$.
\end{theorem}
\emph{Remark}: 
The pseudo-code for the clustering-based adaptive estimation can be found in Algorithm\,\ref{alg:adaptive_estimator}.
The tuning parameters $\lambda_k$ can be thought of as forgetting factors, assigning greater weight to the most recent classification result while retaining the memory of past classification outcomes to enhance robustness.
\begin{algorithm}
\caption{\footnotesize Clustering-Based Adaptive Estimator}
\label{alg:adaptive_estimator}
\footnotesize
\begin{algorithmic}
\Require
\State $N$: Trajectory length or the number of data points on a trajectory
\State $K$: Number of cells in the training set
\State $S^z_n$: Array representing the classification index sequence
\State $Ah_n$: Array representing the accumulated $Ah$ throughput
\State $Q^z_0$: Initial SOH of cell $z$
\State Initialize an array for lambda values with size $K$: $\lambda = [0] * K$
\State Initialize an array for $\bar{Q}^z$ values with size $N$: $\bar{Q}^z = [0] * N$
\Ensure
\State Calculate the denominator for lambda:
\State $\text{denominator} = 0$
\For{$n = 1$ to $N$}
    \State $\text{denominator} += Ah_n[n]$
\EndFor
\State Calculate lambda values for each training cell:
\For{$k = 1$ to $K$}
    \State $\text{numerator} = 0$
    \For{$n = 1$ to $N$}
        \If{$S^z_n[n] = k$}
            \State $\text{numerator} += Ah_n[n]$
        \EndIf
    \EndFor
    \State $\lambda[k]$ = \text{numerator}/\text{denominator}
\EndFor
\State Calculate $\bar{Q}^z$ for the cell in the test set:
\For{$n = 1$ to $N$}
    \State $\bar{Q}^z[n] = 0$
    \For{$k = 1$ to $K$}
        \State $\bar{Q}^z[n] += \lambda[k] \cdot \bar{Q}^k(Ah_n[n])$
    \EndFor
\EndFor
\State Calculate $Q^z$:
\For{$n = 1$ to $N$}
    \State $Q^z[n] = Q^z_0 \cdot \bar{Q}^z[n]$
\EndFor
\State The estimated SOH of the cell in test set is stored in the array $Q^z$
\end{algorithmic}
\end{algorithm}

\begin{proof}
From (\ref{eqn:lambda_adaptation}), $\lambda_k$ is designed to be positive. Moreover, $\lambda_k$ satisfies:
\begin{align} \label{eqn:lambda_sum_property}
    \sum_{k = 1}^K \lambda_k = 1, 
\end{align}
% \begin{align} \label{eqn:lambda_unit_property}
%     0 \le \lambda_k \le 1, \quad \text{for}\;\; 1 \le k \le K. 
% \end{align}
% Property (\ref{eqn:lambda_sum_property}) 
This can be proved by:
\begin{align}
    \sum_{k = 1}^K \lambda_k & = \sum_{k = 1}^K \frac{\sum_{n = 1}^{N} \resizebox{!}{1.5\height}{$\chi$}(S^z_n = k) Ah_n}{\sum_{n = 1}^{N} Ah_n} \\
    & = \sum_{n = 1}^{N} \frac{\sum_{k = 1}^K 
    \resizebox{!}{1.5\height}{$\chi$}(S^z_n = k) Ah_n}{\sum_{n = 1}^{N} Ah_n} \\
    & = \sum_{n = 1}^{N} \frac{Ah_n}{\sum_{n = 1}^{N} Ah_n} = 1.
\end{align}
% Because of (\ref{eqn:lambda_sum_property}) and $\lambda_k$ is positive, Property (\ref{eqn:lambda_sum_property}) can be easily verified.

To prove the BIBO stability, from Definition\,\ref{eqn:bibo_dde}, the $K$ SOH trajectories $ \{Q_n^{1}\}, \cdots, \{Q_n^{K}\}$ in the training set are bounded:
\begin{align} \label{eqn:train_y_bd}
    \|Q^{k}\| < \infty,\quad \quad \text{for}\;\; 1 \le k \le K.
\end{align}
From Definition\,\ref{eqn:bibo_dde}, the SOH trajectory $\{Q^{z}_n\}$ in the test set is bounded:
\begin{align} \label{eqn:test_y_bd}
    \|Q^{z}\| < \infty.
\end{align}
The error trajectory between the estimated capacity and the true capacity follows: 
% \begin{align}
%     \text{RMSE}(\hat{Q}_{adaptive}, Q) \le \epsilon_2 
% \end{align}
\begin{align}
    \big \| \hat{Q}^{z} - Q^{z}\big \| 
    \stackrel{(\ref{eqn:norm_q_z})}{=} & \Bigg \| Q_{0}^z \hat{\bar{Q}}^{z}  - Q^{z} \Bigg \| \\ 
    \stackrel{(\ref{eqn:norm_q_z_decomp})}{=} & \Bigg \| Q_{0}^z\sum_{k = 1}^{K} \lambda_k
    \bar{Q}^k - Q^{z} \Bigg \| \\
    \stackrel{(\ref{eqn:lambda_sum_property})}{=} & \Bigg \| Q_{0}^z\sum_{k = 1}^{K} \lambda_k
    \bar{Q}^k - \sum_{k = 1}^{K} \lambda_k Q^{z} \Bigg \|\\
% \end{align}
% \begin{align}
    \stackrel{(\ref{eqn:norm_q_ch})}{=} & \Bigg \| Q_{0}^z\sum_{k = 1}^{K} \lambda_k
    \bar{Q}^k -   Q_{0}^z \sum_{k = 1}^{K} \lambda_k \bar{Q^{z}} \Bigg \| \\
    = & Q_{0}^z \Bigg \|  \sum_{k = 1}^{K} \lambda_k
    \left(\bar{Q}^k - \bar{Q}^z \right) \Bigg \| \label{eqn:delta_q_bd1}.
\end{align}
From (\ref{eqn:lambda_sum_property}) and (\ref{eqn:delta_q_bd1})
\begin{align}
    \big \| \hat{Q}^{z} - Q^{z}\big \| 
    \le & Q_{0}^z \underset{1\le k \le K}{\text{max}} \{\big \|
    \bar{Q}^k - \bar{Q}^z \big \|\} \\
    \le &  \underset{1\le k \le K}{\text{max}} \{\big \|
    Q^k\big \|\} + \|Q^z\| < \infty.
\end{align}
% The term $\big \|
% \bar{Q}^k_{} - \bar{Q}^z_{} \big \|_{\infty}$ represents the worst-case difference in normalized C/20 capacity within the dataset.
Therefore, the error trajectory is bounded. From Definition \ref{eqn:bibo_dde}, the BIBO stability is guaranteed.
\end{proof}
This proof aims to demonstrate that the estimation error of the test cell's SOH is bounded by a function of the SOH values of all the cells in the training set.
% Finally, the estimated C/20 charge capacity trajectory is obtained from 
% \begin{align}
%     \hat{Q}(Ah_n) = Q_{bsl}^z \bar{\hat{Q}}^k,
% \end{align}
\begin{figure*}[thpb]
\centering
\includegraphics[width=0.8\textwidth]{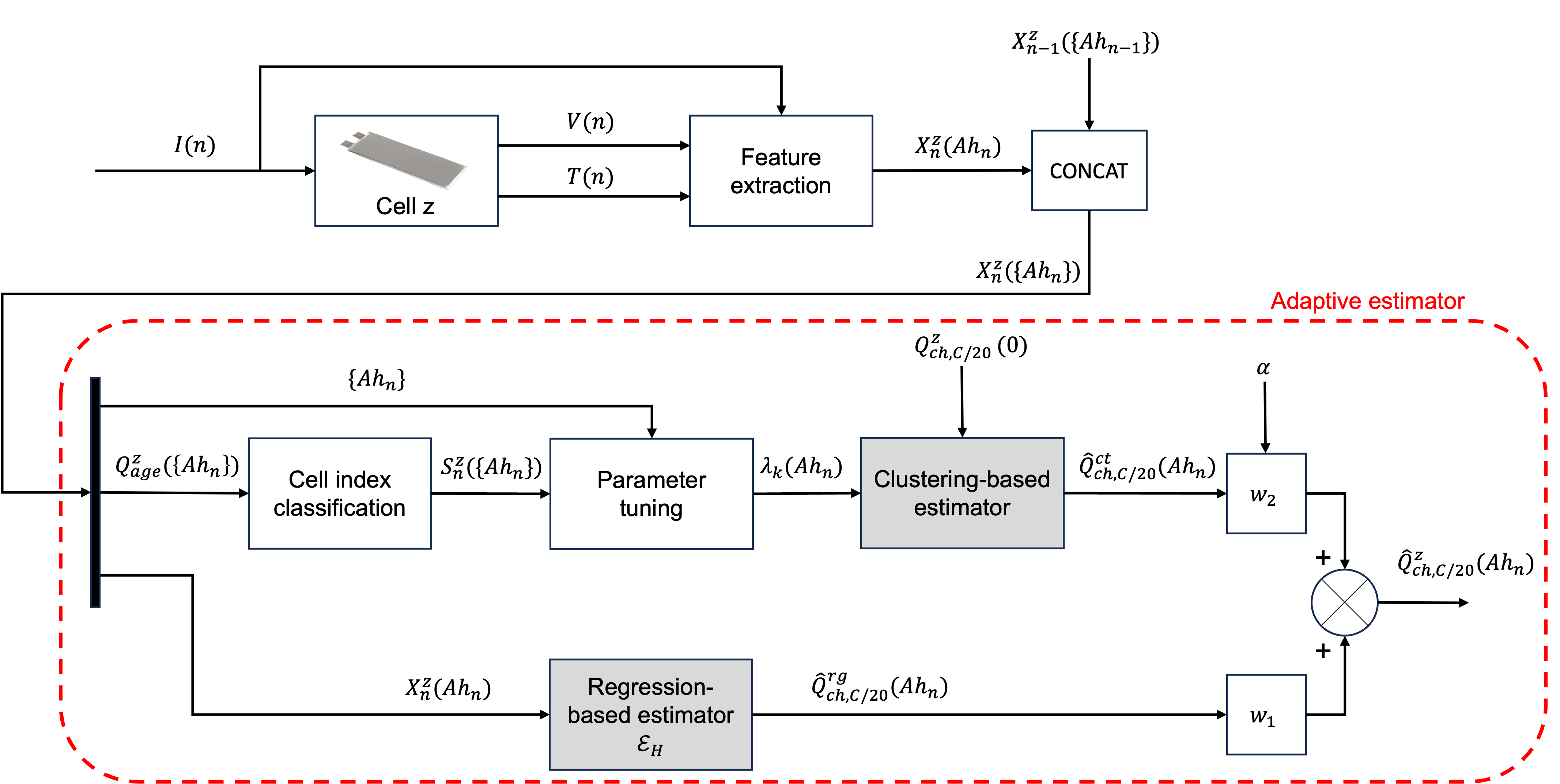}
\caption{\small Block structure showing current $I(n)$, voltage $V(n)$, and cell surface temperature $T(n)$ used to extract features at time $n$. The extracted features are concatenated with features at time instant $n-1$ and this complete sequence of features is input to the adaptive estimator. Output $\hat{Q}_{ch,C/20}^z$ is obtained through a weighted sum of the output from clustering-based estimator and regression-based estimator (offline ENR model). Clustering-based estimator uses only the complete sequence of Ah-throughput $\{Ah_n\}$ and aging cycle charge trajectory $Q^z_{age}(\{Ah_n\})$ that is observed while all the signals from $X_n^z$ at the current time instance are used for the regression-based estimation.}
\label{fig:adaptive_estimator}
\end{figure*}

\subsection{Adaptive Estimation by Combining the Clustering and Regression}
\label{sec:stability_analysis}
The clustering-based method proposed in this study can exhibit a significant classification error when there is limited observation available for the cell under test. To address this limitation, we combinine the clustering-based estimation and the regression-based estimation.
\begin{theorem}
Consider a data-driven estimator described in Definition\,\ref{def:dd_soh_est}. 
%Let the battery cell $z$ be the cell under test.
The adaptive weight parameters $w_1, w_2 \in \mathbb{R^{+}}$ follows
\begin{align}
    w_2 & = \begin{cases}
             \alpha Ah & \text{if } \alpha Ah < 0.5 \\
             0.5 & \text{otherwise}
           \end{cases}\\
    w_1 & = 1 - w_2,
\end{align}
where $\alpha$ is a fixed constant. Then the adaptive estimator, formulated as
\begin{align} \label{eqn:merge_est}
    \hat{Q}^z_{}(Ah_n) = w_1 \hat{Q}^{rg}_{}(Ah_n)  + w_2 \hat{Q}^{ct}_{}(Ah_n),
\end{align}
is BIBO stable.
\end{theorem}
\emph{Remark 1}: The adaptive estimation is obtained by summing the two estimation results with weights assigned to them. The weight parameters $w_1$ and $w_2$ allocate increasing importance to the online model as time progresses, while still preserving the influence of the offline model.

\emph{Remark 2}: $\alpha$ can be interpreted as a learning-rate constant, measured in units of $Ah^{-1}$, with a range of values from $0$ to $+\infty$. It is recommended to initially set $\alpha$ to $\alpha = 1 / 20 Ah_{\text{max}}$ at the outset, where $Ah_{\text{max}}$ represents the largest Ah throughput that the batteries are expected to be cycled during its second life. Subsequently, $\alpha$ can be adjusted using the data in the training set.
For example, the designer can gradually increase $\alpha$ until the point at which the validation error begins to rise. Sensitivity analysis of $\alpha$ for Cell 1.2, 1.4, and 2.3 is shown in Section~\ref{sec:alpha_sensitivity}.
\begin{proof}
% For ease of notation, in this proof, the variable $Q_{}$ is simplified as $Q$.
%BIBO stability guarantees that the worst-case error in health estimation can be bounded. 
The error trajectory between the estimated capacity and the true capacity follows:
\begin{align}
    & \| \hat{Q}^{z} - {Q}^{z} \| \\
    \stackrel{(\ref{eqn:merge_est})}{=} & \Bigg\| \left(w_1 \hat{Q}^{rg} + w_2 \hat{Q}^{rg} \right) - (w_1 + w_2) {Q}^{z}) \Bigg\| \\
    = & \big\| w_1 (\hat{Q}^{rg} - {Q}^{z}) + w_2 (\hat{Q}^{ct} - {Q}^{z}) \big\| \\
    \le &  w_1 \big \| \hat{Q}^{rg} - Q^{z} \big \| + w_2 \big \| \hat{Q}^{ct} - Q^{z} \big \| \label{eqn:th2_stable}
    % \le & \text{max}\big\{\big \|\hat{Q}_{}^{rg} - Q^{z}_{} \big \|, \big \| \hat{Q}_{}^{ct} - Q^{z}_{} \big \|\big\}.
\end{align}
From Theorem\,\ref{theorem:ddm_stable}, the second term $w_2\|\hat{Q}^{ct} - Q^{z}\|$ in (\ref{eqn:th2_stable}) is bounded.
To prove the stability, from Definition \ref{eqn:bibo_dde}, the feature trajectories in the training set are bounded:
\begin{align} \label{eqn:train_x_bd}
    \|X^{train}_{age}\| < \infty.
\end{align}
From (\ref{eqn:train_x_bd}) and (\ref{eqn:test_y_bd}), the first term in (\ref{eqn:th2_stable}) is bounded by
\begin{align}
   & \big \| \hat{Q}^{rg} - Q^{z} \big \| \\
  = &  \big \| X^{train}_{age} \beta + \beta_0 - Q^{z} \big \|\\
  \le & \|X^{train}_{age} \|\|\beta\| + \|\beta_0\| + \|Q^{z}\| < \infty.
\end{align}
Therefore, the error trajectory is bounded and the BIBO stability is guaranteed.
% Since the regression model is affine, and the input features are bounded, the resulting health estimation output is also bounded.
% \begin{align}
%     \text{RMSE}(\hat{Q}_{offline}, Q) \le \epsilon_1  
% \end{align}
\end{proof}

The complete flow of information for the adaptive estimator is shown in Fig.\,\ref{fig:adaptive_estimator}. In summary, the adaptive law used in the estimation process does not cause the estimation to diverge, and the worst-case error in health estimation can be bounded. The size of the error ball is determined by the errors of the offline model and adaptive model. Furthermore, the radius of the error ball can be further reduced by tunning the gains $w$ and $\lambda$.

\subsubsection{Sensitivity analysis of learning-rate constant $\alpha$} \label{sec:alpha_sensitivity}
\begin{figure}[thpb]
\centering
\includegraphics[width=0.8\columnwidth]{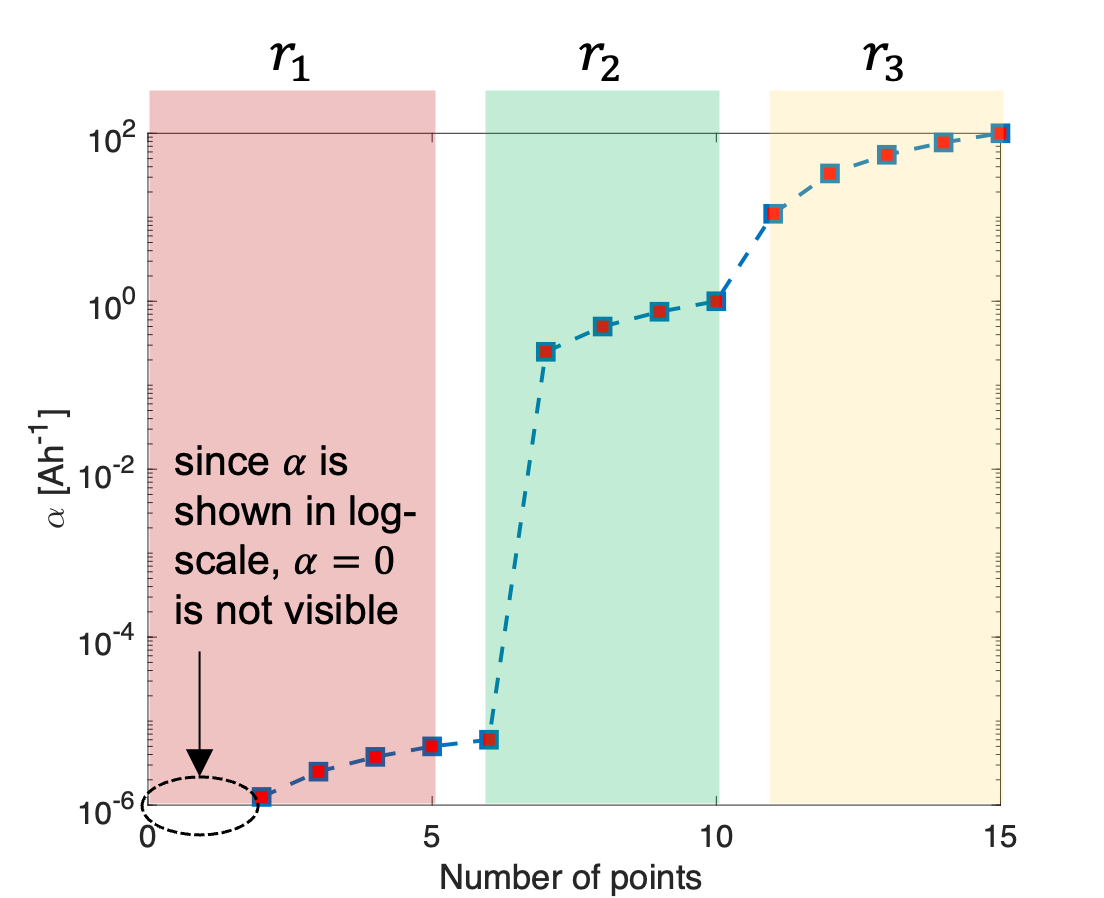}
\caption{\small Range of $\alpha$ values used for sensitivity analysis shown on log-scale.}
\label{fig:alpha_points}
\end{figure}

For the adaptive estimator given in Section~\ref{sec:stability_analysis}, $\alpha$ is used to adapt $w_2$ to control the contribution of $\hat{Q}^{ct}(Ah_n)$ in the output. The range is given by $\alpha = [0, +\infty)$, which means it is important to choose a suitable value of $\alpha$ to ensure the best performance from the estimator. In this sensitivity analysis, three different ranges of $\alpha$ values are tested, and each range consists of five linearly spaced points, including the boundary values. These ranges are given by: 1) Range $r_1 = [0, 5 \times 10^{-6}]$, 2) Range $r_2 = [6 \times 10^{-6}, 1]$, and 3) Range $r_3 = [11, 100]$ as shown in Fig.~\ref{fig:alpha_points}.

\begin{figure}[thpb]
\centering
\includegraphics[width=0.85\columnwidth]{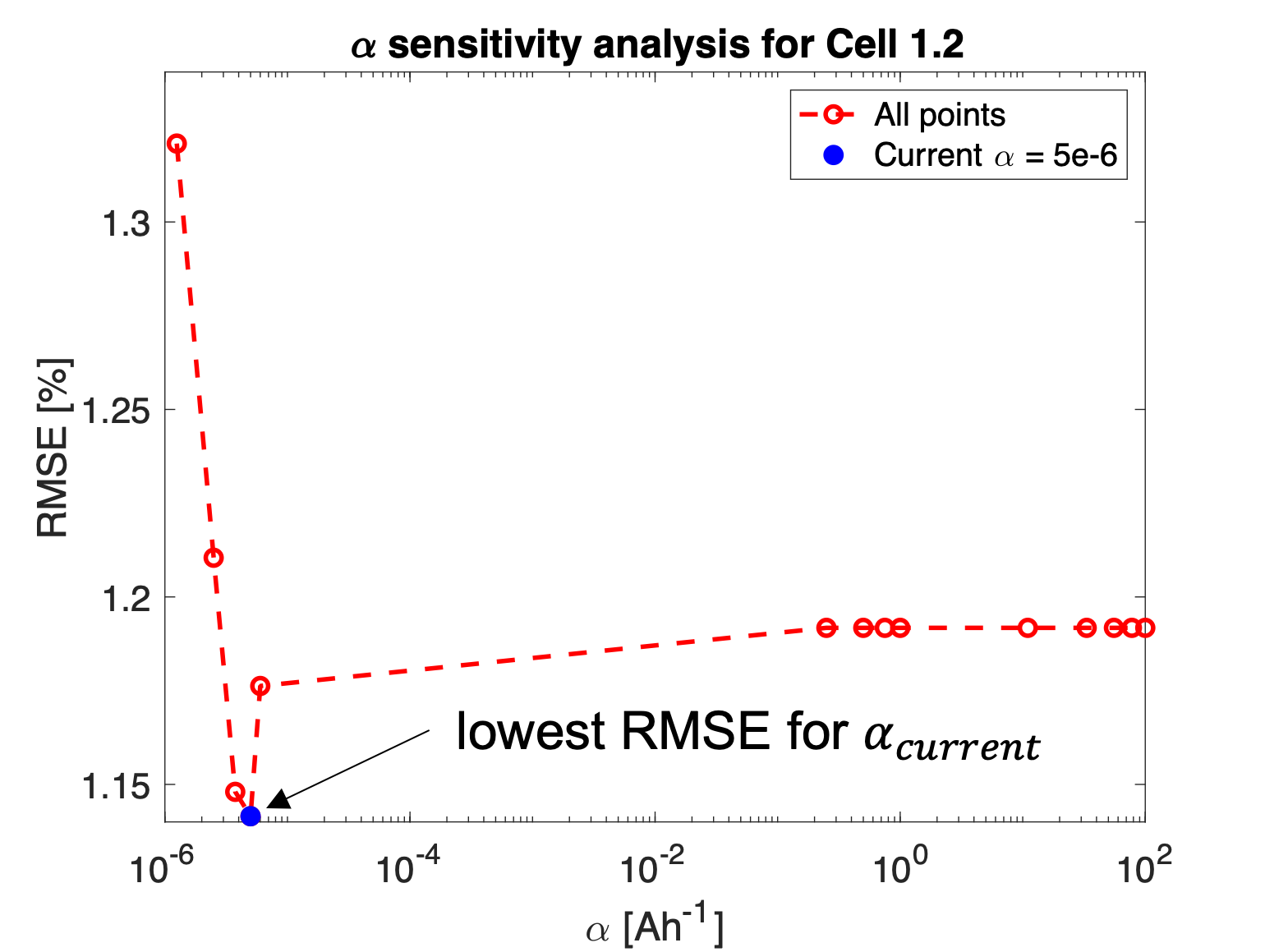}
\caption{\small Sensitivity analysis of $\alpha$ for Cell 1.2 with $\alpha_{current}$ giving the lowest RMSE value}
\label{fig:sensitivity_1p2}
\end{figure}

\begin{figure}[thpb]
\centering
\includegraphics[width=0.85\columnwidth]{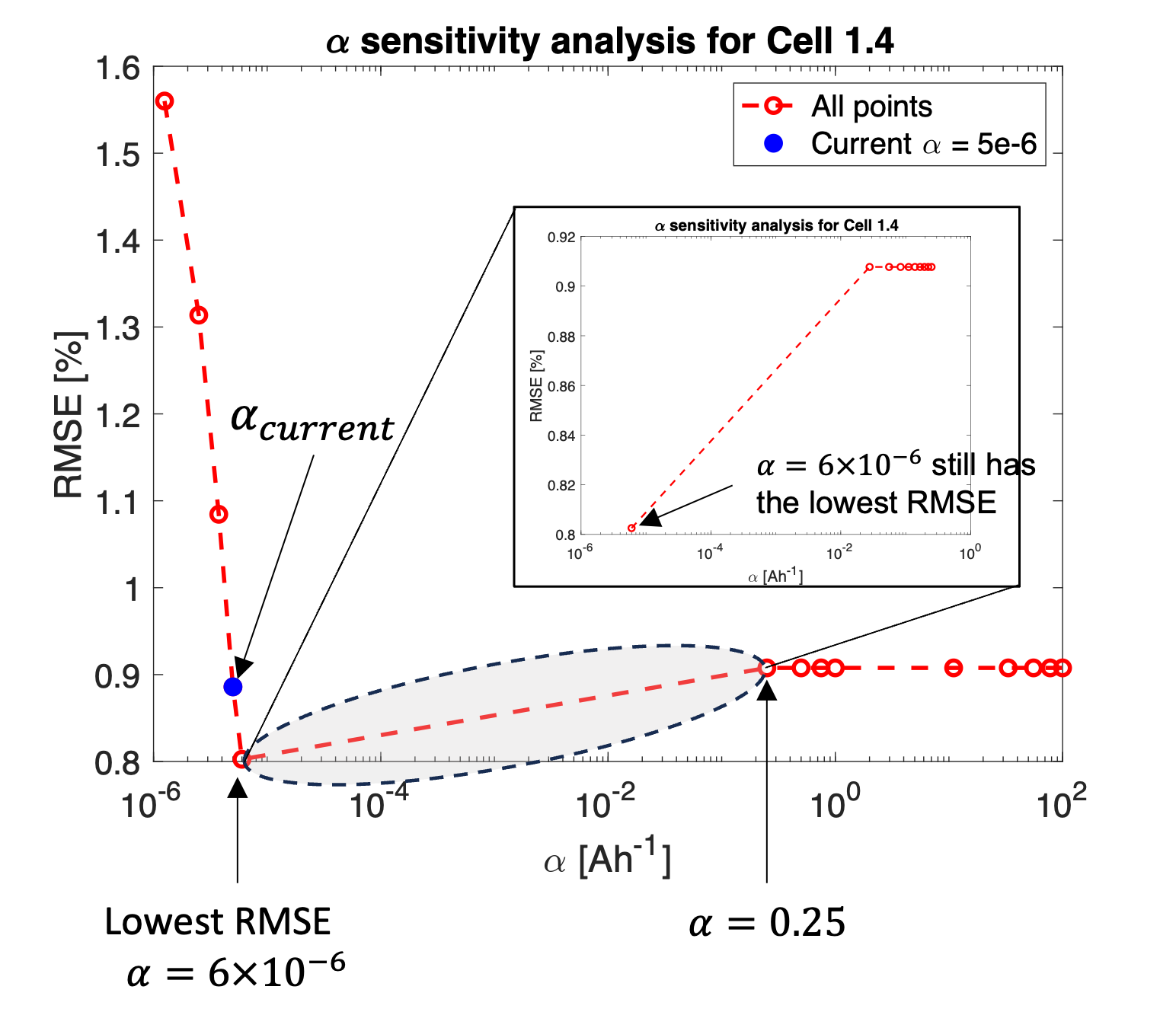}
\caption{\small Sensitivity analysis of $\alpha$ for Cell 1.4 with $\alpha = 6\times10^{-6} > \alpha_{current}$ giving the lowest RMSE value}
\label{fig:sensitivity_1p4}
\end{figure}

For sensitivity analysis, the RMSE error is observed for different $\alpha$ values and the results are compared to the current $\alpha$ value used in our results. Fig.~\ref{fig:sensitivity_1p2} shows the sensitivity of $\alpha$ for cell 1.2. It can be seen that $\alpha_{current}$ -- the current value of $\alpha$ -- gives the lowest RMSE for the entire range of values. In fact, using values smaller or greater than $\alpha_{current}$ increases the RMSE value. Repeating the same analysis for cell 1.4 as shown in Fig.~\ref{fig:sensitivity_1p4}, we can see that $\alpha_{current}$ does not give the smallest RMSE; instead, the smallest RMSE is obtained at $\alpha = 6 \times 10^{-6}$. An extra set of linearly spaced 10 points between $[6 \times 10^{-6}, 0.25]$ is analyzed and it can be seen that lowest RMSE is still obtained at the same value. It should be noted that $\alpha = 5 \times 10^{-6}$ is a close second in terms of lowest RMSE with a difference of less than $0.1\%$ for the lowest RMSE value.

\begin{figure}[thpb]
\centering
\includegraphics[width=0.85\columnwidth]{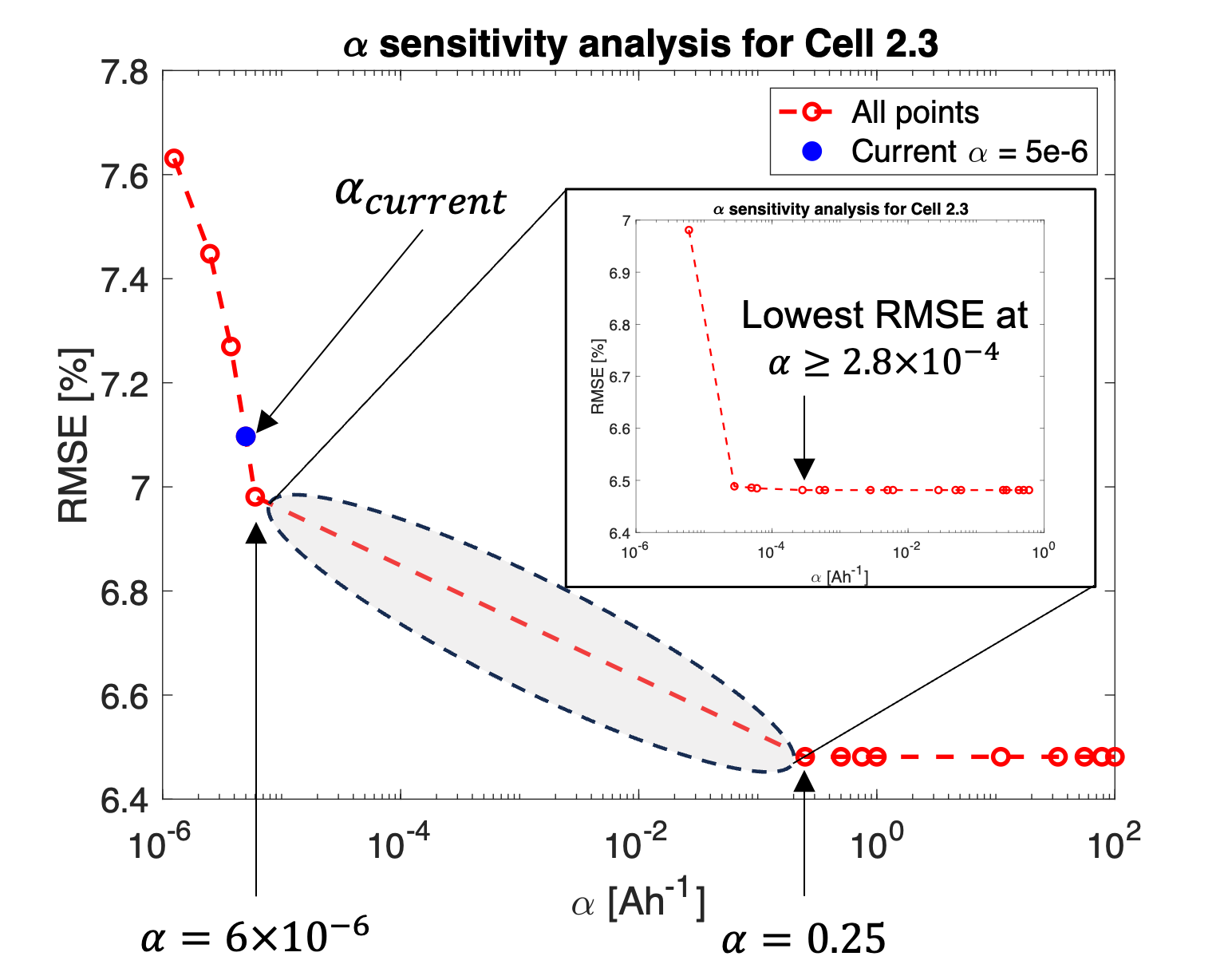}
\caption{\small Sensitivity analysis of $\alpha$ for Cell 2.3 with $\alpha \geq 2.8\times10^{-4} > \alpha_{current}$ giving the lowest RMSE values}
\label{fig:sensitivity_2p3}
\vspace{-10pt}
\end{figure}

Finally, for Cell 2.3, the sensitivity analysis results are shown in Fig.~\ref{fig:sensitivity_2p3}. It can be seen that neither $\alpha = 5 \times 10^{-6}$ or $\alpha = 6 \times 10^{-6}$ give the lowest RMSE value. Instead, by analyzing 18 further points between $[6 \times 10^{-6}, 0.25]$, it can be seen that the lowest RMSE is obtained for a range of values of $\alpha \geq 2.8\times10^{-4}$. One thing to note here is that Cell 2.3 has the largest errors in the offline ENR model which means relativity larger contribution of the clustering-based estimator $\hat{Q}^{ct}(Ah_n)$ is needed to improve the estimation of the model.

\section{Results and Discussion}
\begin{figure}[htbp]
    \centering
    \includegraphics[width = 6.5cm]{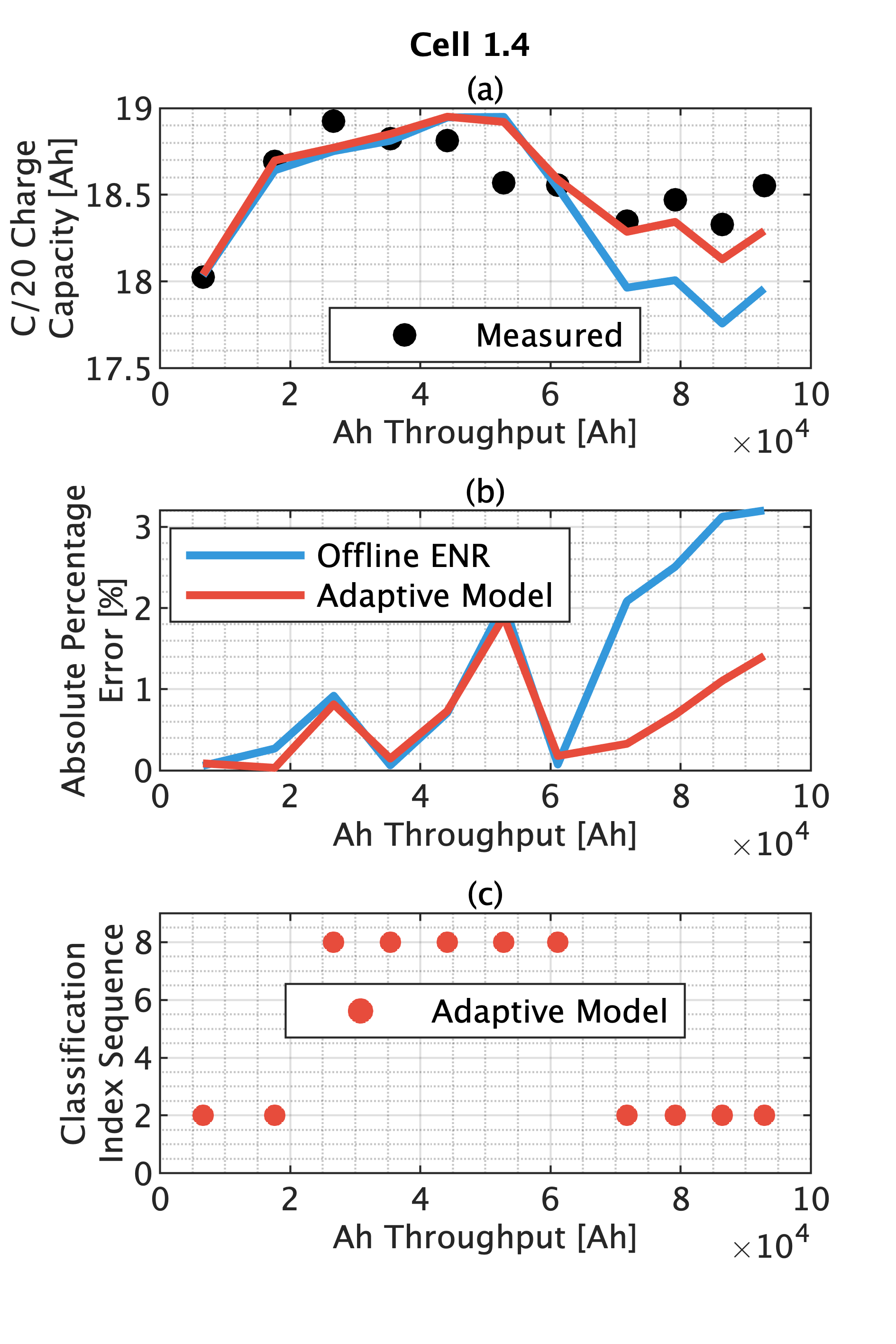}
    \caption{\small Performance evaluation of the online adaptive estimator for cell 1.4. (a) The C/20 charge capacity trajectories estimated from the offline Elastic-Net Regression (ENR) and online adaptive model are compared against the measured capacity. (b) The pointwise absolute capacity estimation percentage error of the offline ENR is compared against that of the online adaptive model. The Root Mean Square Error (RMSE) of the online adaptive model is  0.8610\,\%. (c) The classification index sequence, as defined in Definition\,\ref{eqn:class_ind_seq}, of the online adaptive estimator.}
    \label{fig:Ada_est_cell1p4}
    \vspace{-10pt}
\end{figure}
\begin{figure}[htbp]
    \centering
    \includegraphics[width = 6.5cm]{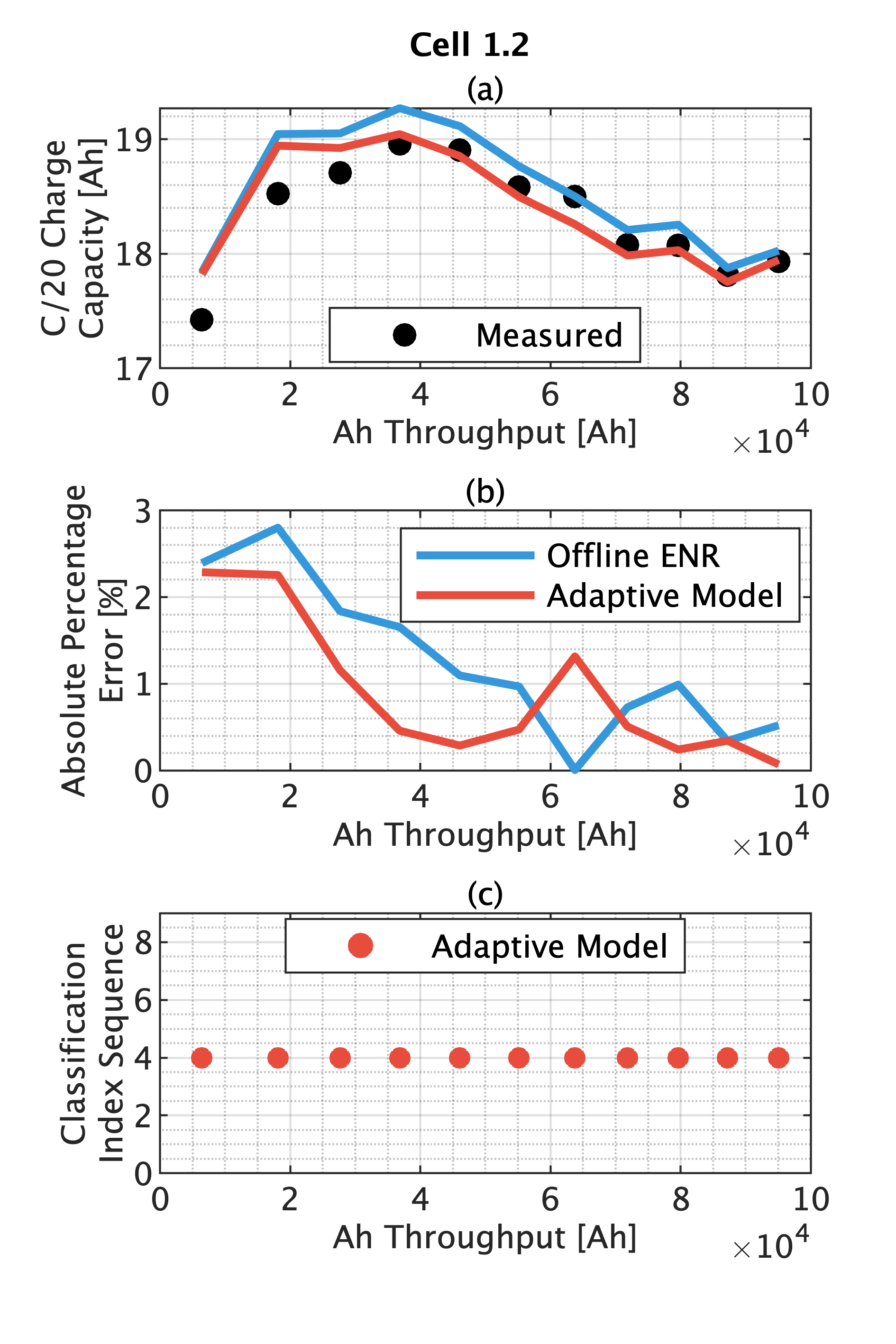}
    \caption{\small Performance evaluation of the online adaptive estimator for cell 1.2. (a) The C/20 charge capacity trajectories estimated from the offline Elastic-Net Regression (ENR) and online adaptive model are compared against the measured capacity. (b) The pointwise absolute capacity estimation percentage error of the offline ENR is compared against that of the online adaptive model. The Root Mean Square Error (RMSE) of the online adaptive model is 1.2182\,\%. (c) The classification index sequence, as defined in Definition\,\ref{eqn:class_ind_seq}, of the online adaptive estimator.}
    \label{fig:Ada_est_cell1p2}
    \vspace{-10pt}
\end{figure}

The performance of the proposed online adaptive estimation algorithm is demonstrated in several case studies. In the first demonstration, we leave out Cell 1.4 for testing while training the model using the remaining seven cells. The C/20 charge capacity trajectories estimated by the ENR model, online adaptive model, and measured capacity are illustrated in Figs.\,\ref{fig:Ada_est_cell1p4}(a), (b), and (c), respectively.
\begin{table*}[htbp] %\label{tab:leave_one_validation}
\centering
\caption{Leave-one-out tests of two health estimation methods.}
\label{tab:leave_one_validation}
\begin{tabular}{|c|c|c|c|c|c|c|c|c|} 
\hline
RMSPE [\%] & Cell 1.1 & Cell 1.2 & Cell 1.3 & Cell 1.4 & Cell 2.1 & Cell 2.2 & Cell 2.3 & Cell 2.4 \\
\hline
Adaptive Estimation & 2.93 & 1.22 & 2.62 & 0.86 & 2.26 & 3.82 & 7.27 & 0.95 \\
\hline
ENR-based Estimation & 2.95 & 1.47 & 2.41 & 1.82 & 1.45 & 3.27 & 7.82 & 1.58 \\
\hline
\end{tabular}
\end{table*}

Initially, when the Ah throughput is lower, the weight $w_1$ is large, while the weight $w_2$ is small. The ENR-based estimate dominates the overall estimate. Therefore, the ENR model and adaptive model coincide at the same starting point, as demonstrated in Fig.\,\ref{fig:Ada_est_cell1p4}(a). As Ah throughput increases to $Ah > 2\times 10^4$ Ah, the classification index converges to the right cluster (cluster 8 / Cell 2.4), as depicted in Fig.\,\ref{fig:Ada_est_cell1p4}(c). However, it is observed that after Ah throughput increases to $Ah > 6.5 \times 10^4$ Ah, the classification index is again perturbed to the wrong cluster (cluster 2 / Cell 1.2). Due to the weight adaptation law (\ref{eqn:lambda_adaptation}), the overall estimation output is only slightly impacted by the wrong classification.

Because the adaptive SOH estimator continuously processes the real-time influx of data, it yields reduced estimation errors. As displayed in Fig.\,\ref{fig:Ada_est_cell1p4}(b), employing solely the ENR model yields a maximum absolute pointwise percentage error of 3\,\%, and a root mean squared pointwise percentage error of 1.8166\,\%. In contrast, by leveraging the online adaptive model that combines the ENR-based estimation and clustering-based estimation, the maximum pointwise percentage error decreases to 2\,\%, and the root mean squared pointwise percentage error decreases to 0.8610\,\%.

This example demonstrates the impacts of weight adaptation. Weight $w$ balances the contributions of the offline ENR model and the online clustering-based model. It avoids the important hazards that the online clustering-based model causes by under-informed classification in the beginning when little data is collected. Weight $\lambda$ balances the contributions of the current-step classification result and historical-steps classification results. These historical-steps classification results are memorized by summing their corresponding Ah throughput. This effectively decreases the high sensitivity of the online clustering-based estimation method to the current-step classification result.

Another demonstration shown in Fig.\,\ref{fig:Ada_est_cell1p2}, the online adaptive estimation algorithm is tested on Cell 1.2, in which the ENR has already illustrated a good performance. As shown in Fig.\,\ref{fig:Ada_est_cell1p2}, adding the online adaptive estimator can further reduce the pointwise percentage RMSE of capacity estimation from 1.4686\,\% to 1.2182\,\%. 

The online adaptive estimation algorithm has also been tested on the other six cells for robust analysis purposes. In each test, we perform a leave-one-out test on the entire dataset, leaving only the cell under test in the test set and including the rest of the seven cells in the training set. The averaged RMSPE over eight test cases of the ENR-based estimation is 3.40\,\%, while the RMSPE over eight test cases of the adaptive estimation is 3.27\,\%. The RMSPE for each training-testing set split is presented in Table\,\ref{tab:leave_one_validation}.
\section{Conclusions}
Due to its high flexibility and model agnosticism, data-driven health estimation has emerged as a valid and viable method for assessing the health of SL batteries.
To enable the in-the-field operation of SL battery energy storage systems, 
% the health estimation method must solely rely on online-accessible operational information from SL batteries and be self-adaptable in real time to accommodate the unique characteristics of each individual cell. Additionally, the adaptation law must be carefully designed while ensuring rigorous stability.
% In this paper, 
we present an online adaptive health data-driven estimation method with guaranteed stability.
The clustering-based estimation is combined together with the elastic-net regression.
We have validated this method using a dataset of lab-aged second-life batteries retired from commercial EVs.
Our method is illustrated to reduce the estimation error compared to the offline health estimation methods.


%%%%%%%%%%%%%%%%%%%%%%%%%%%%%%%%%%%%%%%%%%%%%%%%%%%%%%%%%%%%%%%%%%%%%%%%%%%%%%%%

%%%%%%%%%%%%%%%%%%%%%%%%%%%%%%%%%%%%%%%%%%%%%%%%%%%%%%%%%%%%%%%%%%%%%%%%%%%%%%%%

%%%%%%%%%%%%%%%%%%%%%%%%%%%%%%%%%%%%%%%%%%%%%%%%%%%%%%%%%%%%%%%%%%%%%%%%%%%%%%%%
%\newpage

%\vspace{+10pt}
%$$ \text{ACKNOWLEDGEMENTS} $$

\bibliographystyle{IEEEtran}
\bibliography{acc_main_v12}

\begin{thebibliography}{10}
\providecommand{\url}[1]{#1}
\csname url@samestyle\endcsname
\providecommand{\newblock}{\relax}
\providecommand{\bibinfo}[2]{#2}
\providecommand{\BIBentrySTDinterwordspacing}{\spaceskip=0pt\relax}
\providecommand{\BIBentryALTinterwordstretchfactor}{4}
\providecommand{\BIBentryALTinterwordspacing}{\spaceskip=\fontdimen2\font plus
\BIBentryALTinterwordstretchfactor\fontdimen3\font minus \fontdimen4\font\relax}
\providecommand{\BIBforeignlanguage}[2]{{%
\expandafter\ifx\csname l@#1\endcsname\relax
\typeout{** WARNING: IEEEtran.bst: No hyphenation pattern has been}%
\typeout{** loaded for the language `#1'. Using the pattern for}%
\typeout{** the default language instead.}%
\else
\language=\csname l@#1\endcsname
\fi
#2}}
\providecommand{\BIBdecl}{\relax}
\BIBdecl

\bibitem{lutsey2018power}
N.~Lutsey, M.~Grant, S.~Wappelhorst, and H.~Zhou, ``Power play: How governments are spurring the electric vehicle industry.''\hskip 1em plus 0.5em minus 0.4em\relax ICCT Washington, DC, USA, 2018.

\bibitem{eddy2019recharging}
J.~Eddy, A.~Pfeiffer, and J.~van~de Staaij, ``Recharging economies: The ev-battery manufacturing outlook for europe,'' \emph{McKinsey \& Company}, 2019.

\bibitem{pavlinek2023transition}
P.~Pavl{\'\i}nek, ``Transition of the automotive industry towards electric vehicle production in the east european integrated periphery,'' \emph{Empirica}, vol.~50, no.~1, pp. 35--73, 2023.

\bibitem{jones2023electric}
B.~Jones, V.~Nguyen-Tien, and R.~J. Elliott, ``The electric vehicle revolution: Critical material supply chains, trade and development,'' \emph{The World Economy}, vol.~46, no.~1, pp. 2--26, 2023.

\bibitem{dong2023cost}
Q.~Dong, S.~Liang, J.~Li, H.~C. Kim, W.~Shen, and T.~J. Wallington, ``Cost, energy, and carbon footprint benefits of second-life electric vehicle battery use,'' \emph{iScience}, 2023.

\bibitem{lu2022battery}
J.~Lu, R.~Xiong, J.~Tian, C.~Wang, C.-W. Hsu, N.-T. Tsou, F.~Sun, and J.~Li, ``Battery degradation prediction against uncertain future conditions with recurrent neural network enabled deep learning,'' \emph{Energy Storage Materials}, vol.~50, pp. 139--151, 2022.

\bibitem{anna2023}
A.~Weng, E.~Dufek, and A.~Stefanopoulou, ``Battery passports for promoting electric vehicle resale and repurposing,'' \emph{Joule}, pp. 837–--842, 2023.

\bibitem{Hu2022}
X.~Hu, X.~Deng, F.~Wang, Z.~Deng, X.~Lin, R.~Teodorescu, and M.~G. Pecht, ``{A Review of Second-Life Lithium-Ion Batteries for Stationary Energy Storage Applications},'' \emph{Proceedings of the IEEE}, vol. 110, no.~6, pp. 735--753, jun 2022.

\bibitem{Pozzato2021b}
G.~Pozzato, S.~B. Lee, and S.~Onori, ``{Modeling degradation of Lithium-ion batteries for second-life applications: preliminary results},'' \emph{CCTA 2021 - 5th IEEE Conference on Control Technology and Applications}, pp. 826--831, 2021.

\bibitem{JIANG2018754}
\BIBentryALTinterwordspacing
Y.~Jiang, J.~Jiang, C.~Zhang, W.~Zhang, Y.~Gao, and N.~Li, ``State of health estimation of second-life lifepo4 batteries for energy storage applications,'' \emph{Journal of Cleaner Production}, vol. 205, pp. 754--762, 2018. [Online]. Available: \url{https://www.sciencedirect.com/science/article/pii/S0959652618328725}
\BIBentrySTDinterwordspacing

\bibitem{Wei2018}
J.~Wei, G.~Dong, and Z.~Chen, ``Remaining useful life prediction and state of health diagnosis for lithium-ion batteries using particle filter and support vector regression,'' \emph{IEEE Transactions on Industrial Electronics}, vol.~65, no.~7, pp. 5634--5643, 2018.

\bibitem{Takahashi2023}
\BIBentryALTinterwordspacing
A.~Takahashi, A.~Allam, and S.~Onori, ``{Evaluating the feasibility of batteries for second-life applications using machine learning},'' \emph{iScience}, vol.~26, no.~4, p. 106547, 2023. [Online]. Available: \url{https://doi.org/10.1016/j.isci.2023.106547}
\BIBentrySTDinterwordspacing

\bibitem{Zhang2014}
\BIBentryALTinterwordspacing
C.~Zhang, J.~Jiang, W.~Zhang, Y.~Wang, S.~M. Sharkh, and R.~Xiong, ``A novel data-driven fast capacity estimation of spent electric vehicle lithium-ion batteries,'' \emph{Energies}, vol.~7, no.~12, pp. 8076--8094, 2014. [Online]. Available: \url{https://www.mdpi.com/1996-1073/7/12/8076}
\BIBentrySTDinterwordspacing

\bibitem{Bhatt2021a}
A.~Bhatt, W.~Ongsakul, N.~Madhu, and J.~G. Singh, ``{Machine learning-based approach for useful capacity prediction of second-life batteries employing appropriate input selection},'' \emph{International Journal of Energy Research}, vol.~45, no.~15, pp. 21\,023--21\,049, dec 2021.

\bibitem{li2022uncertainty}
X.~Li, Y.~Dai, Y.~Ge, J.~Liu, Y.~Shan, and L.-Y. Duan, ``Uncertainty modeling for out-of-distribution generalization,'' \emph{arXiv preprint arXiv:2202.03958}, 2022.

\bibitem{zhang2022machine}
Y.~Zhang, T.~Wik, J.~Bergstr{\"o}m, M.~Pecht, and C.~Zou, ``A machine learning-based framework for online prediction of battery ageing trajectory and lifetime using histogram data,'' \emph{Journal of Power Sources}, vol. 526, p. 231110, 2022.

\bibitem{she2021offline}
C.~She, Y.~Li, C.~Zou, T.~Wik, Z.~Wang, and F.~Sun, ``Offline and online blended machine learning for lithium-ion battery health state estimation,'' \emph{IEEE Transactions on Transportation Electrification}, vol.~8, no.~2, pp. 1604--1618, 2021.

\bibitem{zhou2013optimized}
J.~Zhou, D.~Liu, Y.~Peng, and X.~Peng, ``An optimized relevance vector machine with incremental learning strategy for lithium-ion battery remaining useful life estimation,'' in \emph{2013 IEEE International Instrumentation and Measurement Technology Conference (I2MTC)}.\hskip 1em plus 0.5em minus 0.4em\relax IEEE, 2013, pp. 561--565.

\bibitem{XING2013811}
\BIBentryALTinterwordspacing
Y.~Xing, E.~W. Ma, K.-L. Tsui, and M.~Pecht, ``An ensemble model for predicting the remaining useful performance of lithium-ion batteries,'' \emph{Microelectronics Reliability}, vol.~53, no.~6, pp. 811--820, 2013. [Online]. Available: \url{https://www.sciencedirect.com/science/article/pii/S0026271412005227}
\BIBentrySTDinterwordspacing

\bibitem{Zhang2022a}
Y.~Zhang, T.~Wik, J.~Bergstr{\"{o}}m, M.~Pecht, and C.~Zou, ``{A machine learning-based framework for online prediction of battery ageing trajectory and lifetime using histogram data},'' \emph{Journal of Power Sources}, vol. 526, apr 2022.

\bibitem{von2022state}
F.~Von~B{\"u}low and T.~Meisen, ``State of health forecasting of heterogeneous lithium-ion battery types and operation enabled by transfer learning,'' in \emph{PHM Society European Conference}, vol.~7, no.~1, 2022, pp. 490--508.

\bibitem{zhang2023voltage}
S.~Zhang, H.~Zhu, J.~Wu, and Z.~Chen, ``Voltage relaxation-based state-of-health estimation of lithium-ion batteries using convolutional neural networks and transfer learning,'' \emph{Journal of Energy Storage}, vol.~73, p. 108579, 2023.

\bibitem{cui2023}
X.~Cui, M.~A. Khan, G.~Pozzato, R.~Sharma, S.~Singh, and S.~Onori, ``Taking second-life batteries from exhausted to empowered using experiments, data analysis, and health estimation,'' \emph{Cell Reports Physical Science}.

\bibitem{Ha2024}
\BIBentryALTinterwordspacing
S.~Ha, G.~Pozzato, and S.~Onori, ``{Electrochemical characterization tools for lithium-ion batteries},'' \emph{Journal of Solid State Electrochemistry}, vol.~28, no.~3, pp. 1131--1157, 2024. [Online]. Available: \url{https://doi.org/10.1007/s10008-023-05717-1}
\BIBentrySTDinterwordspacing

\bibitem{aiken2022li}
C.~P. Aiken, E.~R. Logan, A.~Eldesoky, H.~Hebecker, J.~Oxner, J.~Harlow, M.~Metzger, and J.~Dahn, ``Li[\ch{Ni_{$0.5$}Mn_{$0.3$}Co_{$0.2$}}] \ce{O2} as a superior alternative to \ce{LiFePO4} for long-lived low voltage li-ion cells,'' \emph{Journal of The Electrochemical Society}, vol. 169, no.~5, p. 050512, 2022.

\bibitem{attia2022knees}
P.~M. Attia, A.~Bills, F.~B. Planella, P.~Dechent, G.~Dos~Reis, M.~Dubarry, P.~Gasper, R.~Gilchrist, S.~Greenbank, D.~Howey \emph{et~al.}, ``“knees” in lithium-ion battery aging trajectories,'' \emph{Journal of The Electrochemical Society}, vol. 169, no.~6, p. 060517, 2022.

\end{thebibliography}

\end{document}